\pdfoutput=1
\documentclass[structabstract]{aa}  
\usepackage{graphicx}
\usepackage{lscape}
\usepackage{longtable}
\usepackage{natbib}
\usepackage{amsmath}
\usepackage{color}
\usepackage{array} 
\usepackage{tikz,array}
\usetikzlibrary{calc}
\newcommand{\tikzmark}[1]{\tikz[remember picture,overlay]\coordinate (#1);}

\newcolumntype{T}[1]{@{\hspace{\tabcolsep}}c@{\hspace{\tabcolsep}\tikzmark{#1}}}

\bibpunct{(}{)}{;}{a}{}{,} 
\usepackage{txfonts}
\begin{document}
  \title{High-resolution imaging of \textit{Kepler} planet host candidates}
  \subtitle{A comprehensive comparison of different techniques}

  \author{J. Lillo-Box
          \inst{1}
          , D. Barrado\inst{1}, H. Bouy\inst{1}          }

  \institute{Departamento de Astrof\'isica, Centro de Astrobiolog\'ia (CSIC-INTA), ESAC campus 28691 Villanueva de la Ca\~nada (Madrid), Spain\\
              \email{Jorge.Lillo@cab.inta-csic.es}}
        

  \date{Accepted for publication in A\&A on April 29, 2014}


  \abstract
  {The \textit{Kepler} mission has discovered thousands of planet candidates. Currently, some of them have already been discarded; more than 200 have been confirmed by follow-up observations { (most by radial velocity and few by other methods), and several hundreds have been validated}. However, the large majority of the candidates are still awaiting for confirmation. Thus, priorities (in terms of the probability of the candidate being a real planet) must be established for subsequent radial velocity observations.}
  {The motivation of this work is to provide a set of isolated (good) host candidates to be further tested by other techniques that allow confirmation of the planet. As a complementary goal, we aim to identify close companions of the candidates that could have contaminated the light curve of the planet host due to the large pixel size of the \textit{Kepler} CCD and its typical PSF of around 6 arcsec. Both goals can also provide  robust statistics about the multiplicity of the \textit{Kepler} hosts. }
  {We used the AstraLux North instrument located at the 2.2 m telescope in the Calar Alto Observatory (Almer\'ia, Spain) to obtain diffraction-limited images of 174 \textit{Kepler} objects of interest. A sample of demoted \textit{Kepler} objects of interest (with rejected planet candidates) is used as a control for comparison of multiplicity statistics. The lucky-imaging technique used in this work is compared to other adaptive optics and speckle imaging observations of \textit{Kepler} planet host candidates. To that end, we define a new parameter, the blended source confidence level (BSC), to assess the probability of an object to have blended non-detected eclipsing binaries capable of producing the detected transit.}
  {We find that 67.2\% of the observed \textit{Kepler} hosts are isolated within our detectability limits, and 32.8\% have at least one visual companion at angular separations below 6 arcsec. Indeed, we find close companions (below 3 arcsec) for the 17.2\% of the sample. The planet properties of this sample of non-isolated hosts are revised according to the presence of such close companions. We report one possible S-type binary  (KOI-3158), where the five planet candidates would orbit one of the components of the system. {We also report three possible false positives (KOIs 1230.01, 3649.01, and 3886.01) due to the presence of close companions that modify candidate properties such that they cannot be considered as planets anymore}. The BSC parameter is calculated for all the isolated targets and compared to both the value prior to any high-resolution image and, when possible, to observations from previous high-spatial resolution surveys in the \textit{Kepler} sample.}
  {}


\authorrunning{J. Lillo-Box et al.}

  \maketitle
%

\section{Introduction}

The {\it Kepler} mission has provided more than 6000 planet candidates\footnote{ Around 3600 candidates have passed all {\it Kepler} requirements. http://exoplanetarchive.ipac.caltech.edu/cgi-bin/ExoTables/nph-exotbls?dataset=cumulative.} ({\it Kepler} objects of interest, KOI) in its more than four years of almost continuous operation. The end of phase K1 operations of the mission (extrasolar planets search) is by contrast the starting point of a new phase, which is the systematic analysis of the immense database produced by the observatory. In particular, the validation of these planet candidates is the first step to obtain a large catalog of confirmed extrasolar planets that help us to understand the formation, properties, evolution and death of planetary systems. More than 200 {\it Kepler} planets have been confirmed so far, which still  represents less than 5\% of the total sample of candidates. Several techniques { (such as radial velocity, light curve variations, or transit timing variations)} have been used to that end. However, the large pixel size of the {\it Kepler} camera (around $4\times 4$ arcsec), the broad point spread function (PSF) of the {\it Kepler} telescope (with a typical\footnote{http://keplerscience.arc.nasa.gov/calibration/KSCI-19033-001.pdf} full width at half maximum of FWHM$\approx 6.4$ arcsec), and the size of the aperture (typically 6-10 arcsec) used to extract the photometry implies the need for obtaining high-resolution images prior to applying these (somehow expensive and time-consuming) confirmation techniques.  { Exhaustive statistical analysis has also provided hundreds of validated planets \citep[e.g. ][]{rowe14, lissauer14}.}

High-resolution imaging observations have been previously carried out in  other planetary samples (apart from {\it Kepler} candidates) with interesting results, such as the cases of WASP-2, TrEs-2 and TrEs-4 \citep[see][]{daemgen09}, where the properties of the confirmed planets were revised due to the presence of close companions detected by the lucky imaging technique. To date, there are four extensive works on the {\it Kepler} sample that use different high-resolution techniques in different wavelength ranges: speckle imaging in the optical range \citep{howell11}, adaptive optics in the near-infrared \citep{adams12,adams13}, adaptive optics in the optical \citep{law13}, and lucky imaging in the optical range \citep[our previous catalog presented in ][]{lillo-box12}. 

These high-spatial resolution surveys are important for three main reasons: 1) ruling out the possibility of chance-aligned sources in specific configurations that could mimic a planetary transit (e.g. background eclipsing binaries), 2) improving the orbital and physical parameters of the transiting object by accounting for possible extra sources in the {\it Kepler} aperture, and 3) detecting possible bound  companions forming S-type binary systems, where the planet orbits one of the components of the system and acts on the other as a gravitational perturber \citep[see ][]{kley10}. These points are crucial in our understanding of planetary systems (formation and evolution) and are the first step of the confirmation of {\it Kepler} planets. Indeed, in some cases, these observations represent a key step in the statistical validation of very small planets (with a mass too low to be detected by current radial velocity instruments) as was the case of Kepler-37b \citep{barclay13}.

In the present work, we release a new sample of lucky imaging observations of {\it Kepler} candidates and  provide the isolated sample of candidates observed in our previous release \citep{lillo-box12}. These isolated {\it Kepler} objects of interest (hereafter KOIs) represent an excellent sample of candidates to be followed-up, given the very low probability of contamination of their {\it Kepler} light curves \citep{barrado13}. In section \S~\ref{sec:observations}, we describe the target sample, the observations and the reduction of the data. The sensitivity limits of the images are shown in section \S~\ref{sec:sensitivity}. In section \S~\ref{sec:results} we provide an update of our previous survey and the sample of isolated KOIs in our entire lucky imaging dataset. Statistics on the number of detected companions are given in section \S~\ref{sec:newresults}. The analysis of the high-resolution images in terms of quality and how they reduce the probability for a particular KOI to have blended eclipsing binaries is explained in section \S~\ref{sec:bsc}. In the case of KOIs with detected close companions, we provide estimations in section \S~\ref{sec:nonisolated} on how the transit depth and the planetary radius are modified due to the presence of such additional sources. A useful and comprehensive comparison between the different high-resolution imaging surveys of {\it Kepler} candidates, using different techniques, is presented in section \S~\ref{sec:comparison}, and  conclusions are summarized in section \S~\ref{sec:conclusions}.

\section{Observations and data reduction \label{sec:observations}}

\subsection{Target selection \label{sec:targets}}

Among the different releases of {\it Kepler} planet host candidates \citep{borucki11,batalha13,burke13}, a sample of few hundred targets were selected to be observed with high spatial resolution imaging. The selection criteria were based on both the interests of the planets themselves and the observational limitation imposed by the instrument/telescope configuration. The latter restriction is given by the combination of the CAHA 2.2m telescope and the AstraLux instrument, which provides detectability limits of $m_{SDSSi}=20-21$ mag in total exposure times of around 2700s. Since we wanted to detect possible companions that are at least 5.0 magnitudes fainter at 1.0 arcsec (fainter visual companions would usually not affect the planet-star properties significantly), the faintest targets that we observed were of $m_{Kep}<18$ mag. From a practical point, except in few exceptions, we avoided observing KOIs that are fainter than $m_{Kep}=16$ mag to ensure the significance of our results.

In total, we observed 230 KOIs (101 KOIs in 2011, already reported in \cite{lillo-box12}, with detailed information about 44 objects with possible companions, 21 KOIs in 2012, and 108 KOIs in 2013) hosting 376 planet candidates. 
Unfortunately, after some of these KOIs were observed, some of their hosted planet candidates were rejected for different reasons (re-analysis of the light curve by {\it Kepler} team, radial velocity observations, etc.). As a consequence, a total of 56 KOIs among our targets (including all but one of the KOIs with $m_{Kep}>16$)  do not seem to host a planet candidate anymore. On the positive side, we have used these 56 demoted KOIs as sample control so that the same study was carried out for this sample. This leaves us with 174 planet host candidates in our sample (97 KOIs in 2011, 20 KOIs in 2012, and 57 KOIs in 2013) by hosting 313 planet candidates. Among these 174 KOIs, nine have all their candidates already confirmed\footnote{As for January 20th, 2014.}: KOI-0041 or Kepler-100, KOI-0069 or Kepler-93, KOI-0082 or Kepler-102, and KOI-1925 or Kepler-409 from \cite{marcy14},  KOI-1529 or Kepler-59 \citep{steffen12},  KOI-0196 or Kepler-41 \citep{quintana13}, KOI-0351 or Kepler-90 \citep{cabrera13}, KOI-0245 or Kepler-37 \citep{barclay13}, KOI-0094 or Kepler-89 \citep{weiss13}, KOI-2133 or Kepler-91 \citep{lillo-box13}, and KOI-0571 or Kepler-186 \citep{quintana14}.

\subsection{Data acquisition and reduction}

We applied the lucky imaging technique to the selected targets to achieve diffraction-limited resolution. We used the AstraLux North instrument located at the 2.2m telescope at the Calar Alto Observatory (Almer\'ia, Spain). The targets were observed along three visibility windows of the {\it Kepler} field during 2011, 2012, and 2013. The results regarding the non-isolated KOIs of observations on 2011 were published in \cite{lillo-box12}. In the present work, we report the results concerning the isolated candidates observed in 2011 and the new results for the 2012-2013 observing runs. 

We used exposure times for the single frames in the range 30-90 milliseconds (which is below the coherence time of the atmospheric turbulence) and set the number of frames accordingly to accomplish our depth requirement (typically 20000-40000 frames). In all cases, we used the full CCD array of the camera ($24\times24$ arcsec). In the same line as in our previous work, this observing configuration ensures the aimed coverage both in contrast and angular separation from the main target.  Table~\ref{tab:observations} lists the observing characteristics (date, individual exposure times, and number of frames) for each target.
 
Data cube images were reduced by using the online pipeline of the instrument \citep[see][]{hormuth07}, which performs basic reduction and selects the highest quality images. Then, it combines the best 1.0\%, 2.5\%, 5.0\%, and 10\% frames with the highest Strehl ratios \citep{strehl1902}. It calculates the shifts between the single frames, performs the stacking, and resamples the final image to have half the pixel size (i.e. around 0.023 arcsec/pixel). { In this paper, we only use the 10\% selection rate images (which we simply call  AstraLux images). We chose this particular selection rate, since it provides the best compromise between a good angular resolution and the largest magnitude depth, according to our previous experience with the instrument and recommendations from Felix Hormuth (PI of the instrument).} 

\subsection{Astrometric calibration \label{sec:astrometry}}

We acquired images of the M15 globular cluster in all three observing seasons to obtain the relative plate solution of the CCD. We used the more than 100 cross-matched sources with the \cite{yanny94} catalog to obtain the plate scale and position angle of the CCD. We compared the angular separations and position angles of more than 1000 randomly selected star pairs in the latter catalog (separations in arcsec) and in our own catalog (separations in pixel units). The derived pixel size and position angle of the CCD for each observing season are shown in Table~\ref{tab:astrometry}. We obtained typical uncertainties of 0.20~mas/px (around 1\% of relative error) for the pixel size. 

\subsection{Source detection and photometry \label{sec:SourceDetection}}

Sources were identified in each image by using our semi-automatic routine that specifically designed for the instrument. The algorithm first detects possible sources in the image whose integrated flux over an aperture of 10 pixels is, at least 3 times greater than the corresponding flux of the sky in the image (measured as the median value of all pixels, assuming that most of the image is not covered by stars). Then, each source candidate is individually checked to fulfill specific criteria, such as having a PSF-like radial profile shape (to reject possible artifacts and cosmic rays) or having magnitudes in the range of 0-30 to reject possible remaining bad pixels. All images were then manually inspected to check the final detected sources.

We then applied aperture photometry to measure the relative magnitudes between objects in the same image. We used the {\it aper} routine in IDL to extract the flux contained within a specified aperture. This aperture is selected for each image by taking the close objects in the field to avoid contamination of close companions into account. Thus, for each image, we have the instrumental magnitudes for all sources and the magnitude differences with respect to the KOI (which we call $\Delta m$). In cases where a close companion (below 3 arcsec) was found, we obtained additional photometry in the $z_{AstraLux}$ filter (equal to the SDSSz filter from the Sloan Digital all-Sky Survey) to characterize the secondary object. 

Absolute calibration was then performed by using the KIC photometry of the KOI and the instrumental magnitudes and magnitude differences of the surrounding objects with respect to the KOI. First, KIC magnitudes were converted to SDSS magnitudes by using the photometric transformations presented by \cite{pinsonneault12} in their equations 3 and 4. According to \cite{brown11}, the KIC images have a full width at half-maximum of 2.5 arcsec. Hence, as stated by the authors, the KIC photometry is unable to resolve the components of close binary stars. According to this, we can consider that their PSF photometry cannot resolve visual companions closer than 2.5 arcsec, so the magnitudes of such KOIs account for the flux of all sources inside such radius. Thus, we can distinguish between two cases to calibrate our photometry:  KOIs with and without companions closer than 2.5 arcsec. 

When any close companion was detected, we derived the photometric zeropoint of our Astralux images using the i-mag provided in the KIC, neglecting atmospheric or instrumental effects.

In the case where a close companion was found, we assume that the KIC magnitude that is converted to the SDSS system ($m_{SDSS}$) is actually the sum of the fluxes coming from all sources inside 2.5 arcsec. If we call $\Delta m^j_{AstraLux}$ to the magnitude difference of the j-th object inside 2.5 arcsec from the KOI, as detected by AstraLux, the calibrated magnitude of the KOIs must read

\begin{equation}
m^{KOI}_{AstraLux,cal} = m_{SDSS}+2.5\log{  \left(  1+\sum_j{10^{0.4 \Delta m^j_{AstraLux}} }   )\right) }.
\end{equation}

Having the calibrated magnitude of the KOI, we can obtain the absolute magnitudes of all companions in the image as $m_{AstraLux, cal}^{C/} = m^{KOI}_{AstraLux,cal}-\Delta m^{C/}_{AstraLux}$, where $C/$ represents the values for the companion . This scheme was then applied to both filters $i$ and $z$ to obtain the absolute SDSS magnitudes of all objects detected in the images. 

\begin{figure}[ht]
   \centering
   \includegraphics[width=0.5\textwidth]{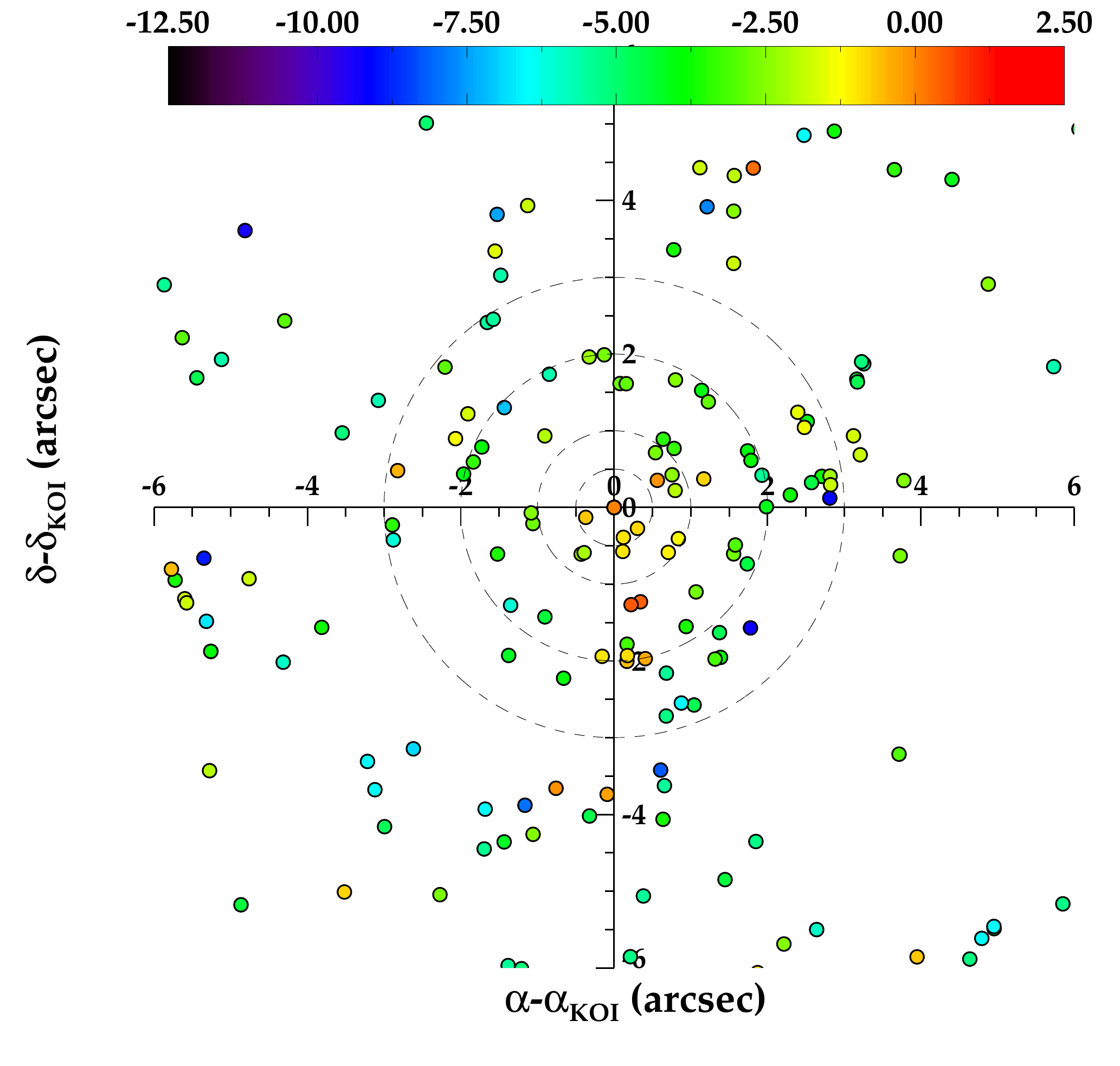}
      \caption{Location of the detected companions to the KOIs in our sample. Each filled circle corresponds to a detected source and its relative position in the projected sky with respect to the KOI. The colors represent the magnitude difference between the companion and the corresponding KOI. We have marked with dashed circles the 0.5, 1.0, 2.0, and 3.0 arcsec separations for visualization purposes. }
         \label{fig:DetectedSources}
   \end{figure}

In Table~\ref{tab:photometry}, we provide the complete catalog of sources detected within 6 arcsec from  the KOIs observed during 2012 and 2013. In Fig.~\ref{fig:DetectedSources}, we show the location of all companions found within 6 arcsec for KOIs that are observed during our three observing seasons. The color-code in the figure shows the magnitude difference in the $i$ filter. Figure~\ref{fig:DetectedSources} illustrates the high density of close visual companions and the need to obtain high resolution images of all candidates to better characterize the systems.

The identified isolated KOIs are studied in more detail in section \S~\ref{sec:bsc}. These targets are thus suitable to proceed with radial velocity studies, since no objects have contaminated the {\it Kepler} light curve and cannot contaminate the radial velocity data within our sensitivity and detectability limits (presented in the next section).

%

\subsection{Completeness, detectability, and sensitivity limits \label{sec:sensitivity}}

In high-spatial resolution studies, it is crucial to determine the limitations of our images in terms of completeness, contrast, and how contrast depends on the separation to the main target. These three concepts completely describe the quality of the observation and, thus, should be individually calculated and reported for each image.


   \begin{figure}[h]
\includegraphics[width=0.5\textwidth]{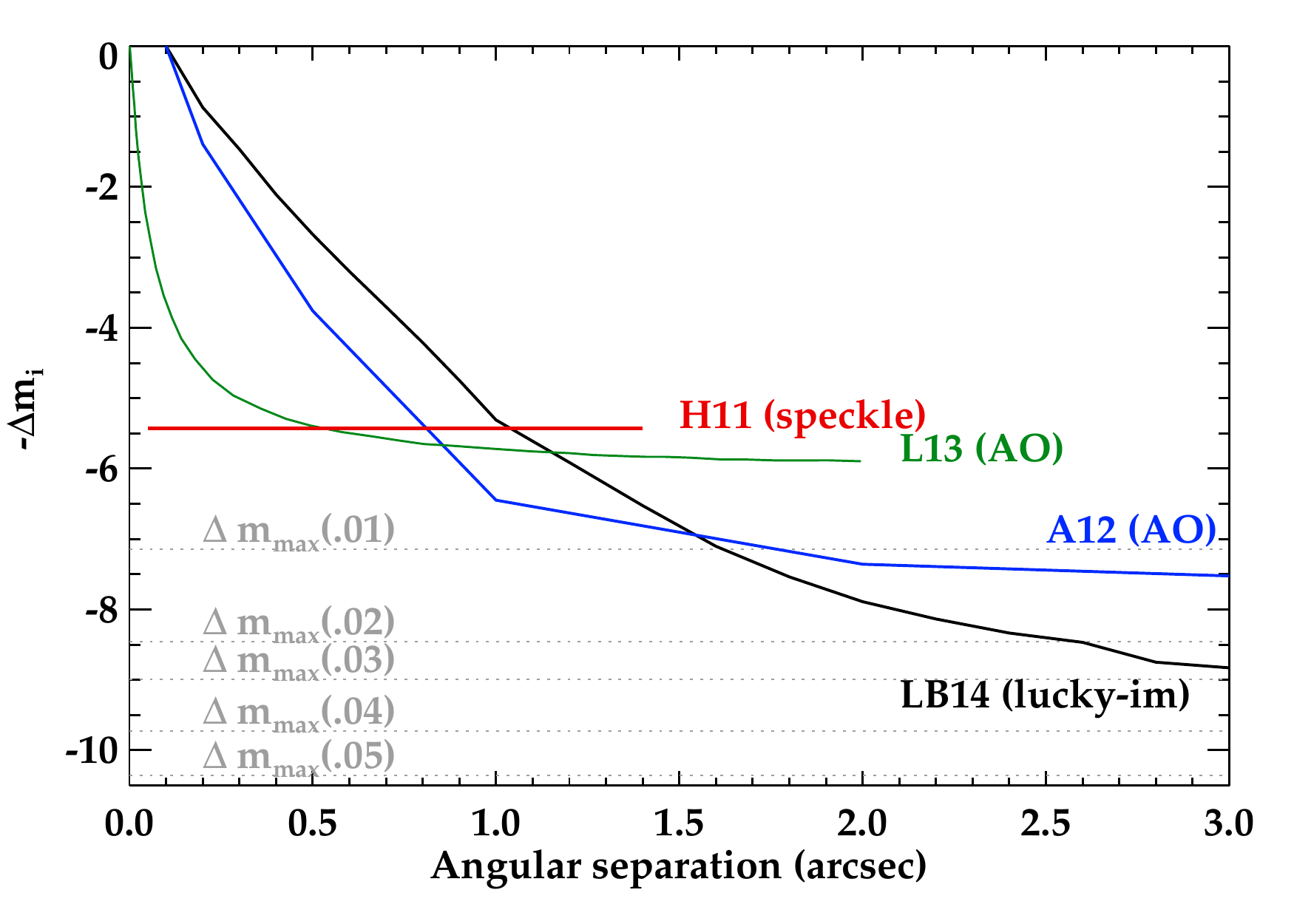}
      \caption{Sensitivity limits ($C_{sens}$) of KOI-0082 in the four main high-resolution imaging surveys in the {\it Kepler} sample, namely \cite{howell11}, red solid line; \cite{adams12}, blue solid line; \cite{law13}, green solid line; and this work, LB14, black solid line. The horizontal grey dotted lines show the maximum magnitude difference that a companion should have to mimic the transit of every planet candidate in that system (see Eq.~\ref{eq:deltam} and section \S~\ref{sec:Ptot})}
         \label{fig:sensitivity}
   \end{figure}

We refer to \cite{lillo-box12} for a detailed explanation about the employed method to measure both the completeness and detectability magnitudes. In brief, we used three images with a total exposure time each of 200s for the globular cluster M15 to compute these parameters. According to the detected sources in these images, we found the mean completeness value to be $i_{complete} = 18.4 \pm 0.3$ mag and to reach detectability down to approximately $i_{detect} = 21.7$ mag {at $5\sigma$ level}. Since different exposure times were set for each image and selection rate, we have to extrapolate these values individually. In particular, we have scaled the completeness and detectability limits for each particular object and selection rate by adding the quantity $-2.5 \log{(200s/t_{exp}(s))}$ to the detectability and completeness magnitudes shown above. Here, $t_{exp}$ is the effective exposure time
of each image (i.e. the individual exposure time per frame multiplied by the number of selected and stacked frames).  The resulting values are shown in Table~\ref{tab:observations} for each individual object.

The sensitivity limits { (i.e. the faintest stars detectable in our images at each angular separation, which we also call sensitivity curve, $C_{sens}$)} were also determined for each image once the KOI was identified. We artificially add a similar PSF compared to the one of the main target but scaled by a factor of $\Delta m$ (i.e. $F_2=F_{KOI} 10^{-0.4\Delta m}$) at different positions of the image. We used 20 angular separations between $\alpha=0.1$ and $\alpha=3.0$ arcsec and 20 relative magnitudes ($\Delta m$) between 0 and 10 magnitudes. For every pair [$d$, $\Delta m$], we added ten artificial stars that are distributed at random angles (hence, 4000 artificial stars were included), and we run our detection algorithm used to detect the companion sources (see section \S~\ref{sec:SourceDetection}) to try detecting these artificially added companions. The sensitivity curve ($C_{sens}$) is then computed as the contour line in the [$d$, $\Delta m$] plane, which shows that at least 7 out of the 10 artificially added stars were detected { with a 5$\sigma$ minimum requirement for the detection}. The sensitivity curve calculated for KOI-0082 is shown in Fig.~\ref{fig:sensitivity} to illustrate the results and to compare other high-spatial resolution surveys to that also observed this target (a quantitative comparison of these studies for the coincident objects is performed in section \S~\ref{sec:comparison}). The sensitivity limits at different angular separations  for each KOI in this work are presented in Table~\ref{tab:sensitivity} { for the corresponding $i$ and $z$ filters.}


\section{Results \label{sec:results}}

In what follows, we use the isolated designation to those KOIs that do not present companions closer than 6 arcsec from the KOI. This limit comes from the typical values of the {\it Kepler} PSF, and the typical apertures used to extract the photometry, which ranges between 6-10 arcsec. Note that stars beyond the 6 arcsec limit would have been detected by more conventional surveys.

\subsection{Update on 2011 results}

In \cite{lillo-box12}, we observed 98 KOIs and concluded that 57 were actually isolated, 27 presented at least one companion between 3-6 arcsec, and 17 presented at least one companion closer than 3 arcsec. The new (more accurate) astrometric calibration presented in this paper with the update of the list of false positives implies that these numbers have slightly changed. The updated results for the 2011 observing season provide 63 isolated KOIs (65.0\%), 22 KOIs with companions at 3-6 arcsec (22.7\%), and 15 KOIs with companions closer than 3 arcsec (15.5\%) out of 97 KOIs (We removed those that currently do not present any planet candidate, namely KOI-0644, KOI-0703, and KOI-0465, and added other two not included in the previous analysis, namely KOI-0490 and KOI-0408.).

\subsection{Results from the new sample \label{sec:newresults}}
In this work, we add information about another 77 KOIs with active planet candidates (nine of them with all their transiting candidates already confirmed, see section \S~\ref{sec:targets}). Among this new sample, we find that 54 KOIs are isolated (70.1\% of the new sample), and 23 KOIs (29.9 \%) have at least one companion closer than 6 arcsec (inside the typical {\it Kepler} PSF). Among the non-isolated KOIs, we find that 12 KOIs (15.6\%) show at least one companion between 3-6 arcsec, and 15 KOIs (19.4\%) show at least one object below 3 arcsec (which means that 4 KOIs present companions in both ranges). These numbers are relatively similar and consistent to those obtained in the 2011 run. 

According to the updated numbers from the 2011 observing season and to the addition of the new sample, we find that 111 are isolated (67.2\%), 35 present at least one companion between 3-6 arcsec (20.1\%), and  30 have companions closer than 3 arcsec (17.2\%), among the 174 KOIs observed with remaining planet candidates,. All these values are summarized in Table~\ref{tab:results}. These results imply that we have a 67.2\% probability that a KOI is isolated, regardless of the possible biases related to the target selection. However, there is a non-negligible 32.8\% probability that a companion object inside the typical {\it Kepler} PSF exists, thus contaminating the {\it Kepler} light curve and modifying the derived properties of any planet candidate.

\section{Analysis \label{sec:analysis}}

\subsection{Isolated planet host candidates \label{sec:bsc}}

Once the sensitivity limits (i.e. the sensitivity curve) have been
calculated for the images of the isolated KOIs (section \S~\ref{sec:sensitivity}), the relevant information to be supplied by our high-resolution observations is how well we can be assured that no blended  background or foreground sources can contaminate the detected transit signal. 

{
Generally, our high-resolution images can play an important role in the rejection of two false positive scenarios. Other configurations, such as hierarchical triples or grazing eclipses cannot be ruled out by high-spatial resolution images, but their occurrence probabilities are extremely low. First and most critical, specific configurations of a blended unassociated eclipsing binary can reproduce the detected planetary transit of the candidate. According to  \cite{morton11} (hereafter MJ11), this case is of particular importance for shallow transits (with apparent depths below $10^3$ ppm), which should be the case of smaller planets. In section \S~\ref{sec:EBlimits}, we deeply investigate and quantify how our high-resolution images can reduce the probability of such scenario being the responsible of the detected transit.

Secondly, even if the transiting object is actually eclipsing the target star, the mere presence of single blended sources not accounted for in the light curve analysis can importantly dilute the transit depth. As a consequence, the transiting object would seem smaller than it actually is. This scenario is discussed and quantified in section \S~\ref{sec:diluted}.
}


\subsubsection{Blended source confidence (BSC): Rejection of background eclipsing binary scenarios \label{sec:EBlimits}}

As stated in MJ11, the probability of having a blended eclipsing binary ($P_{BB}$) inside the PSF of
the KOI can be split into two factors: the probability of having a
blended source ($P_{\rm BS}$) and the probability for that source to be an eclipsing
binary with the appropriate configuration to produce the observed
transit ($P_{\rm appEB}$). While the second factor only depends on
the galactic latitude and magnitude of the KOI (see equation
[14] in MJ11), the first factor can be
observationally constrained to some extent with high-resolution
images. 

To assess the contamination probability, we define a new parameter, the blended source confidence
level (BSC), as the observational level of confidence that no blended
sources are located within a given angular separation to the host
candidate. The BSC is evaluated as the complementary probability of having a  blended object that could mimic the observed transit.

\paragraph{\label{sec:Ptot} \it i.- Probability of having a blended source \vspace{0.3cm} \\}

For a given KOI, we can calculate the maximum contrast (with respect to the measured flux) that a hypothetical blended EB must have
to mimic a transit of fractional depth $\delta$ (see equation [7] in MJ11). This equation reads

\begin{equation}
\label{eq:deltam}
\Delta m_{max}^{kep}=m_{EB}-m_{target}=-2.5\ \log_{10}{(\delta)}.
\end{equation}

This value is valid for the {\it Kepler} filter. Since we are working with
the i band of the SDSS, we have to compute
the magnitude conversion. If we use the KIC ({\it Kepler} Input Catalog)
magnitudes, we can easily see that the {\it Kepler} and $i_{SDSS}$
magnitudes are linearly correlated. Since the vast majority of our
isolated candidates lie in the range $13.0<m_K<16.0$ (only 2 have
$m_K<13.0$), we have only used KOIs in this range to compute the linear
coefficients. We obtain that

$$i_{SDSS}=0.947\cdot m_K + 0.510.$$

The linear correlation goodness of this fit is $R^2=0.98$, which is
acceptable enough for this work. Thus, we can estimate the constrast in the $i_{SDSS}$ band 
as $\Delta i_{SDSS}=0.947\cdot \Delta m_K$, so that

\begin{equation}
\Delta m_{max}^{i_{SDSS}}=0.947\cdot[-2.5\ \log_{10}{(\delta)}].
\end{equation}

\begin{figure}[h!]
   \centering
 \includegraphics[width=0.5\textwidth]{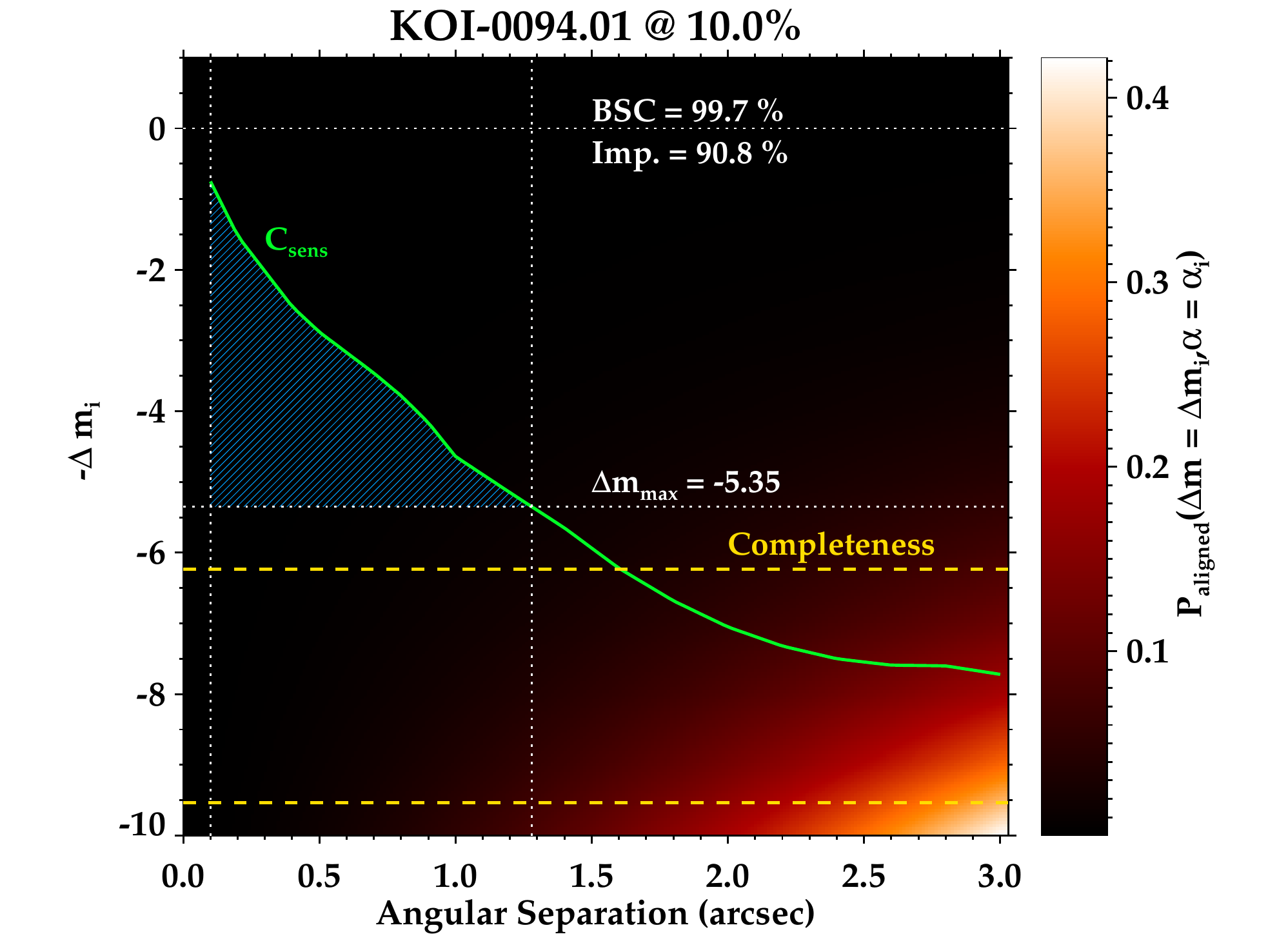}
\vspace{0.3cm}

\includegraphics[width=0.5\textwidth]{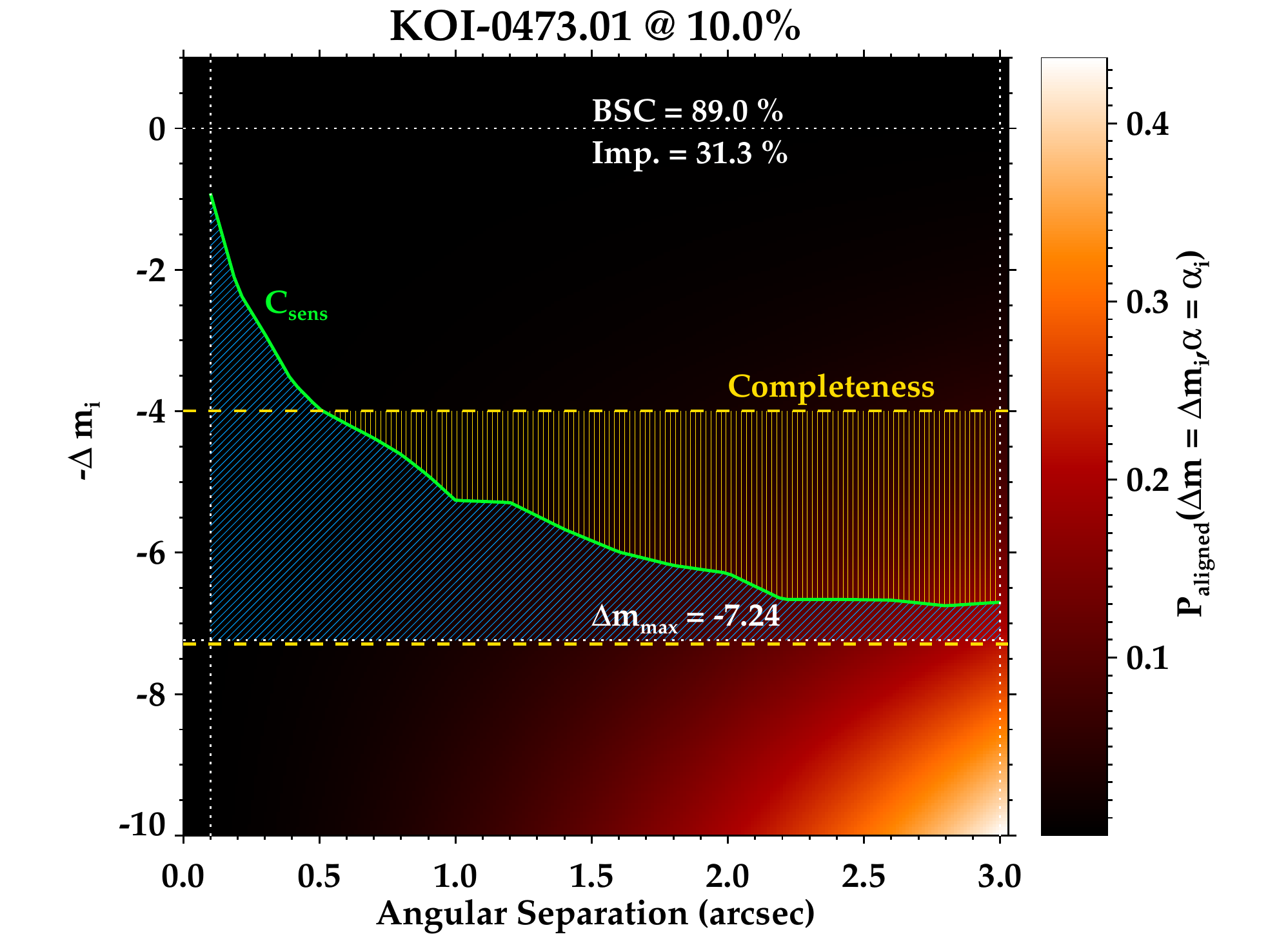}
      \caption{Example of the determination of the BSC parameter (see section \S~\ref{sec:EBlimits}) for two KOIs with conclusive (top panel) and non-conclusive (bottom panel) results. The green solid line represents the 5$\sigma$ sensitivity limit (or sensitivity curve, $C_{sens}$) for the CAHA/AstraLux image (calculated as explained in section \S~\ref{sec:sensitivity}). The lower horizontal dotted white line represents the maximum magnitude difference $\Delta m_{max}$ of a possible eclipsing binary that could mimic the transit signal as expected for these KOIs (see section \S~\ref{sec:EBlimits}). The two vertical white dotted lines show the lowest angular separation detectable in the image (left line) and the intersection between the sensitivity curve and the $\Delta m_{max}$ (right line). { The upper and bottom dashed yellow lines represent  the completeness and detectability levels (respectively) for the given KOI} . The incompleteness region is marked in the bottom panel by the dashed region with vertical yellow lines. The uncovered region is shadowed with diagonal light blue lines. The background color code in the image represents the probability of having a chance-aligned background source for every angular separation and magnitude difference for the given KOI. The first example corresponds to excellent data (deep enough so that no incomplete region is present), whereas the second example is not deep enough. The $\Delta m=0$ is marked by the upper horizontal dotted line.}
         \label{fig:EBlimits}
   \end{figure}

For clarity, we refer to this maximum constrast in the $i_{SDSS}$ band as $\Delta
m_{max}$. In Fig.~\ref{fig:EBlimits}, we show the sensitivity curves of
two KOIs and the $\Delta m_{max}$ line that marks the limit to be reached by observations that minimize the probability of existence of a non-detected blended source capable of mimicking the transit signal.

{
Let us now, as  a first step, assume one particular KOI with a magnitude $m_i$ (in the SDSS photometric system) at galactic latitude $b$. The expected number of stars within an angular separation $r$ from our KOI and with magnitudes in the range $[m_i, \Delta m_{max}]$ is given by

\begin{equation}
\label{eq:n}
\begin{split}
N(r, b, m_i,\Delta m_{max}) & = \int_{0\arcsec}^r 2\pi\alpha \rho(b, m_i,\Delta m_{max}) d\alpha \\
                                  & =   \pi r^2 \rho(b, m_i,\Delta m_{max}).
\end{split}
\end{equation}

\noindent where $\rho(b, m, \Delta m)$ represents the number of stars per unit area (density of stars) and depends on the galactic latitude ($b$) of the target and the requested magnitude range ($[m, m+\Delta m]$). For small areas\footnote{{ With small area, we mean that it must be accomplished that $r<R_{max}$ where  $R_{max}$ is the radius that provides a value of unity for Eq.~\ref{eq:n}, in that it accomplishes $ \int_0^{R_{max}} 2\pi\alpha \rho(b, m,\Delta m) d\alpha = 1$. We note that all studied KOIs accomplish $R_{max}>3.0$ arcsec.}}, this value can be interpreted as the probability for this area that contains one chance-aligned star within this magnitude range. In that case, we can define the probability of an object having one companion source within a certain magnitude range as $P_{\rm aligned}=N(r, b, m,\Delta m)$.} This equation clearly shows that the probability of a chance-aligned source decays with the square of the angular separation as we approach to the star. Contamination sources  above 3.0 arcsec could have been easily detected by photometric observations as dedicated to the {\it Kepler} field \citep[such as in ][{ or the UKIRT J-band survey observed and supplied by Phil Lucas\footnote{See http://keplergo.arc.nasa.gov/ToolsUKIRT.shtml}}]{brown11},  or by photocenter centroid analysis of the {\it Kepler} images \citep[see, for example, ][]{batalha10}. Hence, in this work, we only take the 0.0-3.0 arcsec region into account. 

{ We can now integrate $P_{\rm aligned}$ in the parameter space $\alpha=[0.0,3.0]$ arcsec and $\Delta m=[0, \Delta m_{max}]$  to compute the total {\it a priori} probability ($P_{BS,0}$) for a particular target of magnitude $m_K$ that contains a chance-aligned source with magnitude $m_i<m<m_i+\Delta m_{max}$ within 3 arcsec:
 
\begin{equation}
\label{eq:ptot}
\begin{split}
P_{BS,0}  & =\int_{0\arcsec}^{3\arcsec} 2\pi \alpha \rho(b,m_i, \Delta m_{max}) \ d\alpha = \\
            & = 9\pi\rho(b,m_i,\Delta m_{max}).
\end{split}
\end{equation}

}

\paragraph{ \it ii.- Calculating the $\rho$ parameter \vspace{0.3cm} \\} 

We have calculated the density of stars parameter ($\rho$) for each planet candidate in a similar
way as MJ11. We used the online tool
TRILEGAL\footnote{http://stev.oapd.inaf.it/cgi-bin/trilegal} to
compute the number of expected stars  with a limiting magnitude in a
particular region of the sky. We used the default
parameters for the bulge, halo, thin/thick discs, and the lognormal
IMF (Initial Mass Function) from \cite{chabrier01}. As the {\it Kepler} field is relatively large and encompasses around 12 degrees in each direction { (and almost 20 degrees in galactic latitude), we have found important differences in the stellar density from galactic latitudes that are close to the galactic disk to those farther from it. Due to the large number of targets in this paper and given that it is not possible to perform an automatic query in TRILEGAL (the user must proceed object by object), { it is not possible to obtain individual populations for each target. Since the $\rho$ parameter just depends on the galactic latitude, we obtained instead stellar populations for regions of 1 deg$^2$ centered at $b=6^{\circ}$ to $b=22^{\circ}$ in steps of $1^{\circ}$ and a galactic longitude fixed to the center of the {\it Kepler} field (i.e. $76^{\circ}$), as seen in  Fig. \ref{fig:KeplerField}.  We then simulate stars up to a magnitude limit of $i_{SDSS}=28$ inside each region according to the galactic model.}


   \begin{figure}[hbp]
   \centering
   \includegraphics[width=0.5\textwidth]{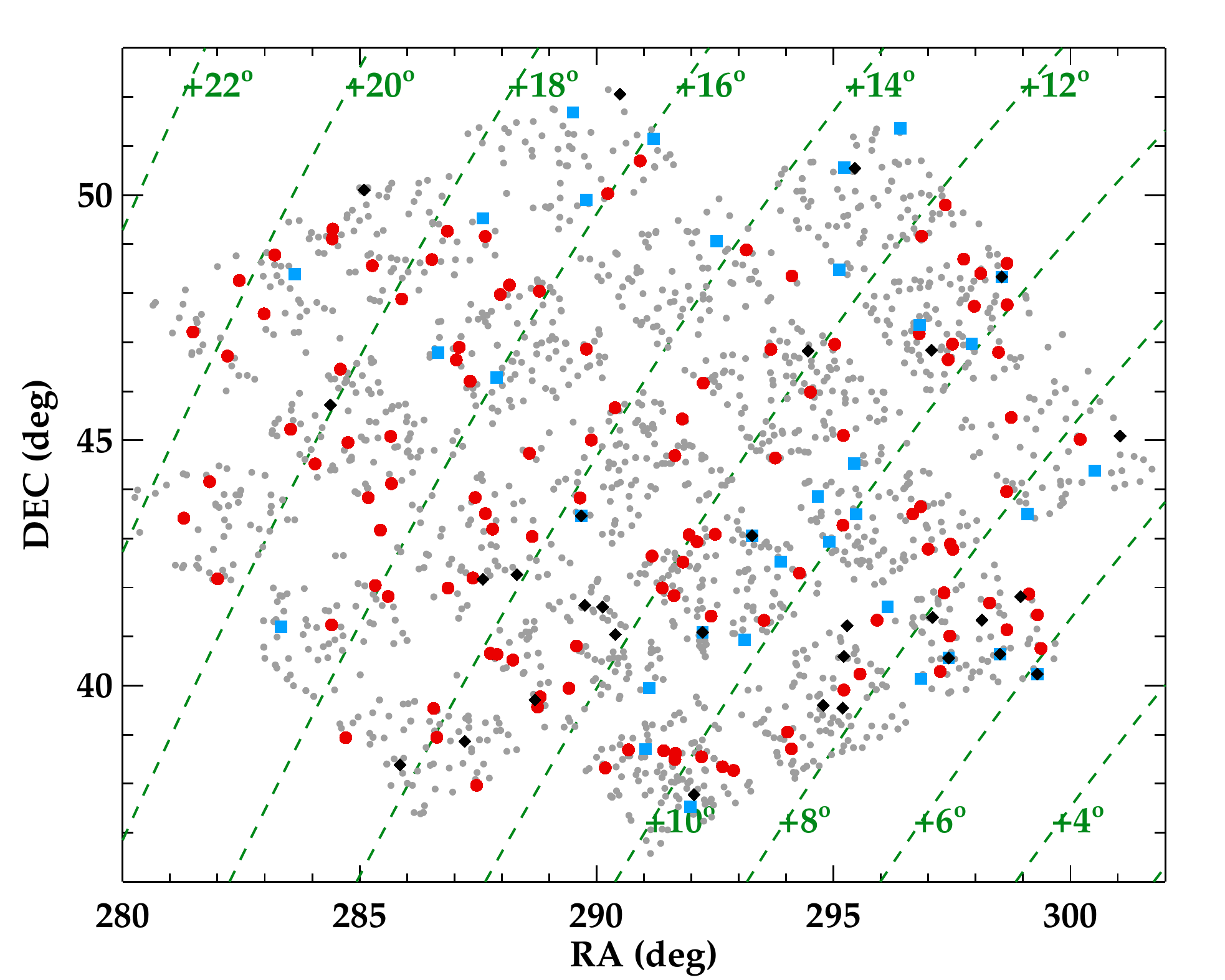}
      \caption{Location of the isolated KOIs (red circles), KOIs with companions at 3-6 arcsec (blue squares), and KOIs with detected companions closer than 3 arcsec (black diamonds), as detected by the CAHA/AstraLux survey. Planet candidates  from \cite{batalha13} are plotted as grey small circles. Iso-galactic latitude lines are marked as dashed inclined green lines parallels to the galactic plane. }
         \label{fig:KeplerField}
   \end{figure}

{ For a particular KOI at a galactic latitude $b_{KOI}$ with a magnitude $m_i$ and a needed depth in magnitude of $\Delta m_i$, we determine $\rho(b_j,m_i,\Delta m_i)$ (j subscript representing { each galactic latitude for which we run the TRILEGAL simulations}) at all galactic latitudes in the grid by just counting stars within the magnitude range $[m_i,m_i+\Delta m_i]$ and dividing by the box area of 1 deg$^2$. Then we perform a low-order polynomial fit to the $rho$ versus galactic latitude values and infer the corresponding $\rho(b_{KOI},m_i,\Delta m_i)$ by evaluating the fitted function at $b_{KOI}$. We found that a polynimial of order five fits the data sufficiently well for the purposes of this work (see the example on Fig.~\ref{fig:GettingRho}). By following this scheme, we can precisely estimate the density of stars in a concrete magnitude range at any position in the {\it Kepler} field.}

   \begin{figure}[htbp]
   \centering
   \includegraphics[width=0.5\textwidth]{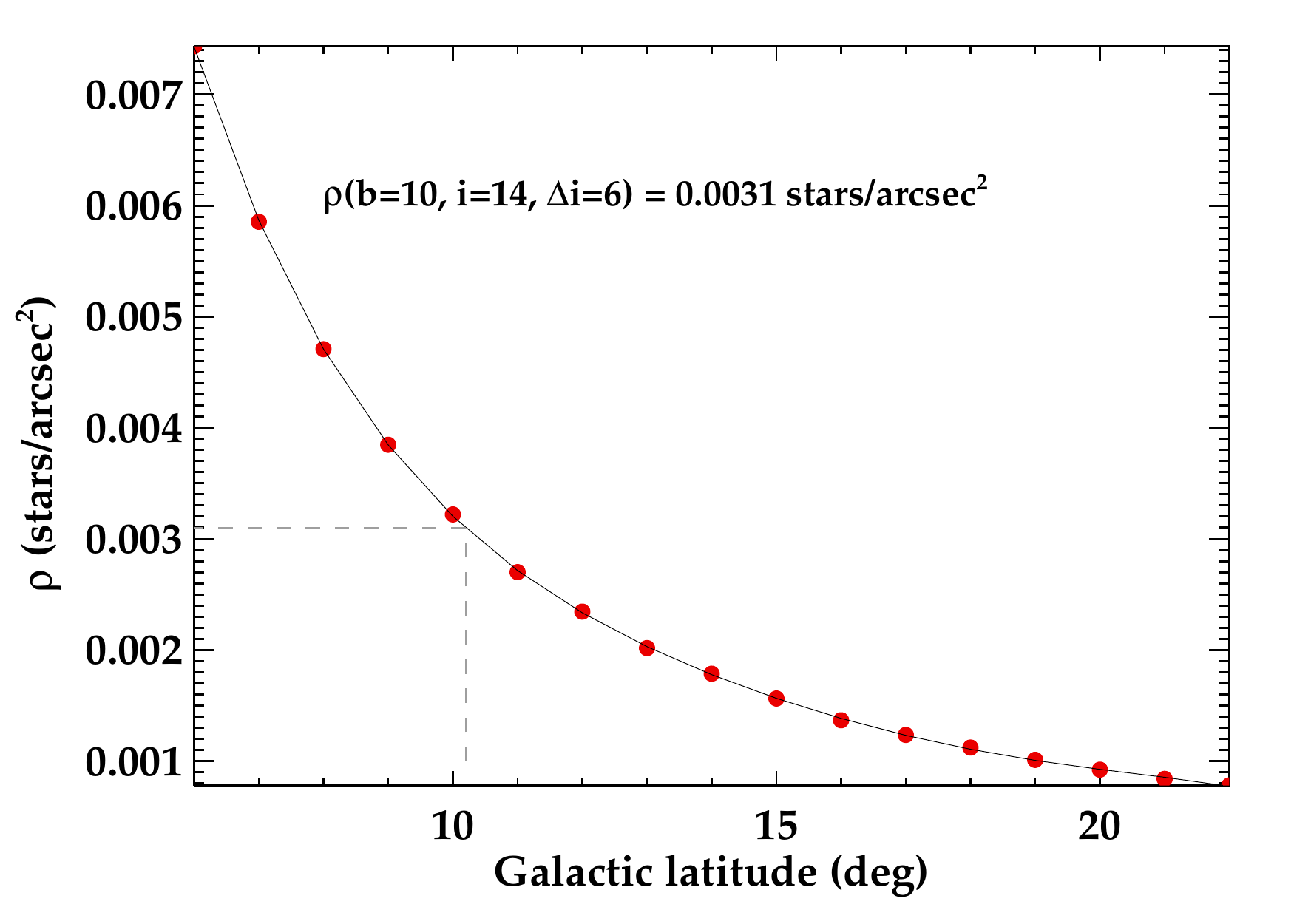}
      \caption{{ Example of the determination of the $\rho$ parameter for an object at $b=10$ deg with a required magnitude range of $i=14-20$ mag. Red filled circles represent the values for each galactic latitude in Fig.~\ref{fig:KeplerField} (i.e. the $\rho[b,m_i,\Delta m_i]$ { with $b$ ranging from $6^{\circ}$ to $22^{\circ}$}), and the solid black line shows the correspoding  five-degree polynomial fit. Gray dashed lines show the derived value at the requested galactic latitude in the example.}}
         \label{fig:GettingRho}
   \end{figure}

\paragraph{ \it iii.- Observing constraints and the BSC parameter \vspace{0.3cm} \\} 

In section \S~\ref{sec:Ptot}, we have described how we define the
probability of having a blended source within some observational
constraints (namely, the star position and the magnitude range of
possible blended stars). Since we have calculated a sensitivity
line for each observation, we can reduce the \textit{a priori}
probability ($P_{BS,0}$) by limiting this calculation to only that region where
our image is not sensitive. { In the angular separation versus magnitude difference plane, this non-sensitive zone is defined as the region  between our sensitivity curve ($C_{sens}$) and the maximum magnitude difference of a possible blended eclipsing binary that could mimic our transit signal ($\Delta m_{max}$)}. In Fig.~\ref{fig:EBlimits}, this region has been shaded with diagonal lines. Hence, with the CAHA/AstraLux high-resolution observations, the probability is given by Eq.~\ref{eq:ptot} but now integrates only over the diagonally shaded region:

{ 

\begin{equation}
\label{eq:pBS}
P_{BS}  = P_{BS,0}- \int_{0\arcsec}^{3\arcsec}   2\pi \alpha \times \rho[b,m_i, \Delta m_0(\alpha)] \ d\alpha,
\end{equation}

\noindent where $\Delta m_0(\alpha) = max[C_{sens}(\alpha), \Delta m_{max}]$ and the second term in the right hand side of the expression represents the contribution of the high-resolution image.

}

 It is thus clear that the better and deeper our image { (i.e. the closer $C_{sens}$ is to $\Delta m_{max}$)}, the more we
diminish the blended source probability and thus improve the
planetary candidacy. We can now determine an observational value of $P_{BS}$ and define the BSC parameter as
${\rm BSC}=1-P_{BS}$, representing the confidence for this source by not having blended eclipsing binaries mimicking the planetary transit. We propose this parameter to be used in all high-resolution studies to compare the
different surveys with the adaptive optics, speckle, or
lucky-imaging techniques. In column six of Table~\ref{tab:bsc}, we show the
updated values of the $P_{BS}$ according to our AstraLux observations (hereafter called $P_{BS}^{LB14}$ in \%) for the isolated targets.

As an example, we obtain {${\rm BSC}= 99.5\%$ in the upper panel of Fig.~\ref{fig:EBlimits}, { given our high-resolution} image of the planet host KOI-0094 and for its KOI-0094.01 planet candidate. In other words, there is a small probability
(lesser than 0.5\%) that we are missing a background source
with the right magnitude to mimic the planetary transit ($P_{BS}<0.5\%$). It must be
remembered that the $P_{BS}$ probability must be multiplied by the
probability of that source being an appropriate eclipsing binary
($P_{appEB}$). This value, as calculated with the correlation
explained in equation [14] in MJ11, is shown for each planet in the
eighth column of Table~\ref{tab:bsc}

\paragraph{ \it iv.- Corrections to the BSC due to incompleteness \vspace{0.3cm} \\} 

In the bottom panel of Fig.~\ref{fig:EBlimits}, we can see an
example of an image providing poor quality information. It is the case of KOI-0473. In this case, we
have to make a correction to the BSC value because the
completeness limit of our observation (corresponding to $i_{complete}$, see section \S~\ref{sec:sensitivity}) is above the sensitivity curve. Since the
detectability limit ($i_{detect}$) is below the sensitivity curve (i.e. $i_{KOI}+\Delta m_{max} < i_{detect}$), we still could detect a
percentage of sources with magnitudes above the completeness magnitude but not the all of them. Thus the contribution
of the vertically shaded region (which we call incompleteness region, Incomp.Reg. in Eq.~\ref{eq:incomp}) must be weighted by a function of the
magnitude difference according to the decay in the probability
detection. This function can be calculated by fitting the decay of the distribution of detected sources in the M15 images to
an exponential function like $f(\Delta m)=C+Ae^{-\Delta m/B}$, where
$\Delta m=m-m_{complete}$. We can
appropriately set $C=0$ and $A=1$, so that $f(\Delta m=0.0)=1.0$. From
the M15 data (see section \S~\ref{sec:sensitivity}), we derived the 
value  $B=0.667$. Thus, in the cases where the completeness line (upper yellow dashed line in Fig.~\ref{fig:EBlimits}) lies
above the { sensitivity curve ($C_{sens}$, green line)}, the $P_{BS}$ must be calculated as

\begin{equation}
\begin{split}
\label{eq:incomp}
P_{BS}^{corr}  =  P_{BS,0}  & -\int_{r_0}^{3\arcsec} 2\pi \alpha \rho[b,m_i, \Delta m_1(\alpha)] \ d\alpha \ \ +  \\
                                   & + \int_{r_0}^{3\arcsec} \ d\alpha \int_{0}^{\Delta' m(\alpha)} 2\pi \alpha \rho[b,m_{comp}, \Delta m_2(\alpha)] \ e^{\frac{-\Delta m_2(\alpha)}{B}}  \ d\Delta'm,
\end{split}
\end{equation}


{ \noindent where 
\begin{align}
\Delta m_1(\alpha) &= max[m_i+C_{sens}(\alpha),m_i+\Delta m_{max},m_{comp}]-m_i \\
\Delta m_2(\alpha) &= max[m_i+C_{sens}(\alpha),m_i+\Delta m_{max}]-m_{comp}
\end{align}

\noindent and $r_0$ represents the angular separation at which $m_i+C_{sens}(\alpha=r_0)=m_{comp}$. The second term corresponds to the contribution of the high-resolution image in the magnitude range where it is complete (non-shaded region above the $C_{sens}$ line in Fig.~\ref{fig:EBlimits}). The third term represents the weightned contribution of the high-resolution image according to our exponential incompleteness function derived above, and it is represented by the vertically shaded region in the bottom panel of Fig.~\ref{fig:EBlimits}.

}

The results for KOI-0473.01 are worse than for KOI-0094.01, since we obtain a BSC parameter of 84.9\% here, indicating that the depth of the image is less than what was needed for this particular object.

We have calculated the $P_{BS}^{LB14}$ parameter for all KOIs without companions closer than 3 arcsec. The results are presented in column 6 of Table~\ref{tab:bsc}. In this table, we also show the $P_{BS,0}$ value (column 5) and the corresponding {\it improvement} obtained with our high-resolution observations (column 7).  These values allow us to compare the quality of different high-resolution imaging techniques.

\subsubsection{Rejecting diluted single-star scenarios \label{sec:diluted}}

{
The second scenario that our high-resolution images can rule out is the case in which the presence of single-blended stars can dilute the transit depth of the eclipsing object so that it could mimic a planetary eclipse. Let us start with the simple case of one single star blended in the {\it Kepler} aperture. The observed transit depth can be calculated as $\delta = (F_{nt}-F_t)/F_{nt}$, where $F_{nt}$ represents the measured flux when the object is not passing in front of the target star and $F_t$ represents the flux when it is transiting the star. In the case of one single star contributing with a flux $F_2$ and the target star contributing with a flux $F_1$, we have $F_{nt} = F_1 + F_2$ and $F_t = F_1(1-\epsilon )$, where $\epsilon$ is the actual fraction of the star covered by the transiting object and its value can be easily demonstrated to be equal to the true transit depth (i.e. $\epsilon =\delta_{true}$). By using this, we can correct the transit depth due to the presence of a blended source as

\begin{equation}
\delta^{true} = \delta^{obs} \left( 1+10^{-\Delta m/2.5} \right)^{1/2},
\end{equation}

\noindent where $\Delta m=m_2-m_1$ represents the magnitude difference between the the blended ($m_2$) and the target star ($m_1$) in the {\it Kepler} band. At this point, as stated by \cite{law13}, we can distinguish between two cases: A) The transited star is brighter than the blended star ($\Delta m>0$), and B) the transited star is fainter than the blended star ($\Delta m<0$). To get the true radius of the transiting object, case B requires some knowledge about the radius ratio between the two stars involved. This requires additional knowledge of both stars (which means more free parameters rather than just the magnitude difference), which is out of the scope of this theoretical analysis. 

In case A, however, assuming that the transit depth is releated to the radius ratio between the transiting object and the parent star as $\delta=(R_p/R_{\star})^2$,  the true radius of the transiting object is given by

\begin{equation}
R_p^{true} = R_p^{obs} \left( 1+10^{-\Delta m/2.5} \right)^{1/2}.
\end{equation}

According to the most updated catalog of confirmed exoplanets\footnote{{ We have checked the radii of the radial velocity confirmed extrasolar planets provided by The Extrasolar Planet Encyclopaedia (http://exoplanet.eu).}}, the empirical maximum possible radius for a planet is $R_p^{max}\approx 2.2~R_J$.  Thus, we can calculate the maximum magnitude difference $\Delta m_{max}^{dil}$ that the blended source can have (i.e. how faint could it be) to dilute the transit depth, such that a non-planetary object (i.e. $R_p>R_p^{max}$, regardless of its nature) appears as the true planet-size object:

\begin{equation}
\label{eq:diluter}
\Delta m_{max}^{dil}= -2.5\log \left[  \left(\frac{R_p^{max}}{R_p^{obs}}\right)^2-1 \right].
\end{equation}

This equation indicates that the presence of undetected blended objects with magnitudes $m_1<m_2<m_1+\Delta m_{max}^{dil}$ can dilute the transit depth of a non-planetary object down to that of a planetary object. According to this expression, case A only applies to candidates with  $R_p^{obs}> 1.56~R_J$. In our sample, we only have six objects with $1.56~R_J < R_p^{obs}< 2.20~R_J$ (namely, KOI-0338.01, 1353.01, 1452.01, 2481.01, 3728.01, and 3765.01 ). For those cases, we can proceed exactly as we did for the blended eclipsing binary scenario to get the $P_{BS}$, but now we use $\Delta m_{max}^{dil}$ as the maximum magnitude value to get the  probability of the presence of a diluter source $P_{DS}$. The results show that this probability is dimminished from $P_{DS,0}=10^{-3}-10^{-2}$ to $P_{DS}^{LB14}=10^{-5}-10^{-4}$. Although the starting probabilities were already small, our high-resolution images showing no blended sources within our detection limits practically discard this possibility as a false positive scenario for these candidates.

We note that due to the mathematical shape of Eq.~\ref{eq:diluter}, we cannot perform this calculation for planet candidates with $R_p^{obs}>R_p^{max}$ (36 in our sample). Also, as stated by \cite{law13}, case B would only affect few planet candidates with observed radii close to the limit $R_p^{obs} \approx 1.56~R_J$ and present blended stars with very small radius. Candidates with a small calculated radius are not affected by this scenario, although the presence of blended sources can modify their properties as we show in section \S~\ref{sec:EffectProperties}. 

For cases with more than one blended star, the $P_{DS}$ probability is insignificant for case A, since the probability of having two or more undeteced sources within our sensitivity limits is far smaller. 

}

\subsection{Non-isolated KOIs \label{sec:nonisolated}}
We have found close companions (below 3 arcsec) for 15 KOIs among the sample of targets in the 2012 and 2013 observing seasons (see Fig.~\ref{fig:closecompanions}). The mere presence of such objects affects both the KOI status as a planet candidate and (if confirmed by other techniques) the derived planet properties, such as planet radius or impact parameter. Thus, the light curves of these targets should be studied in more detail, taking this additional sources into account. In Table~\ref{tab:photometry}, we provide the list of KOIs with companions below 6 arcsec. Among them, 15 KOIs present close companions (below 3 arcsec) that should be added to the 17 KOIs with detected sources by \cite{lillo-box12}. 

Among the 15 KOIs with close companions, we obtained photometry in the $z_{SDSS}$ band for five of them. In these cases, we can compute the $i-z$ color as

\begin{equation}
(i-z)_{/C} = \Delta i - \Delta z + (i-z)_{KOI},
\end{equation} 

\noindent where the subscript $/C$ denotes that the value corresponds to the companion source and $\Delta i =i_{/C}-i_{KOI}$ and $\Delta z =z_{/C}-z_{KOI}$. By following the prescriptions described in \cite{lillo-box12}, we can estimate the spectral types of the objects by having the $i-z$ color information. For those without the $z$ magnitude (mostly KOIs with companions at 3-6 arcsec), we searched for photometric information in public catalogs by using the last version of the Virtual Observatory SED Analyzer \citep{bayo08, bayo13, bayo14}. In cases where more than five photometric points are found, this tool fits a SED model to obtain the effective temperature of the source. Table~\ref{tab:photometry} (column 10) shows the results for the sources for which this study was possible. 

   \begin{figure*}[ht]
   \centering \includegraphics[width=1.0\textwidth]{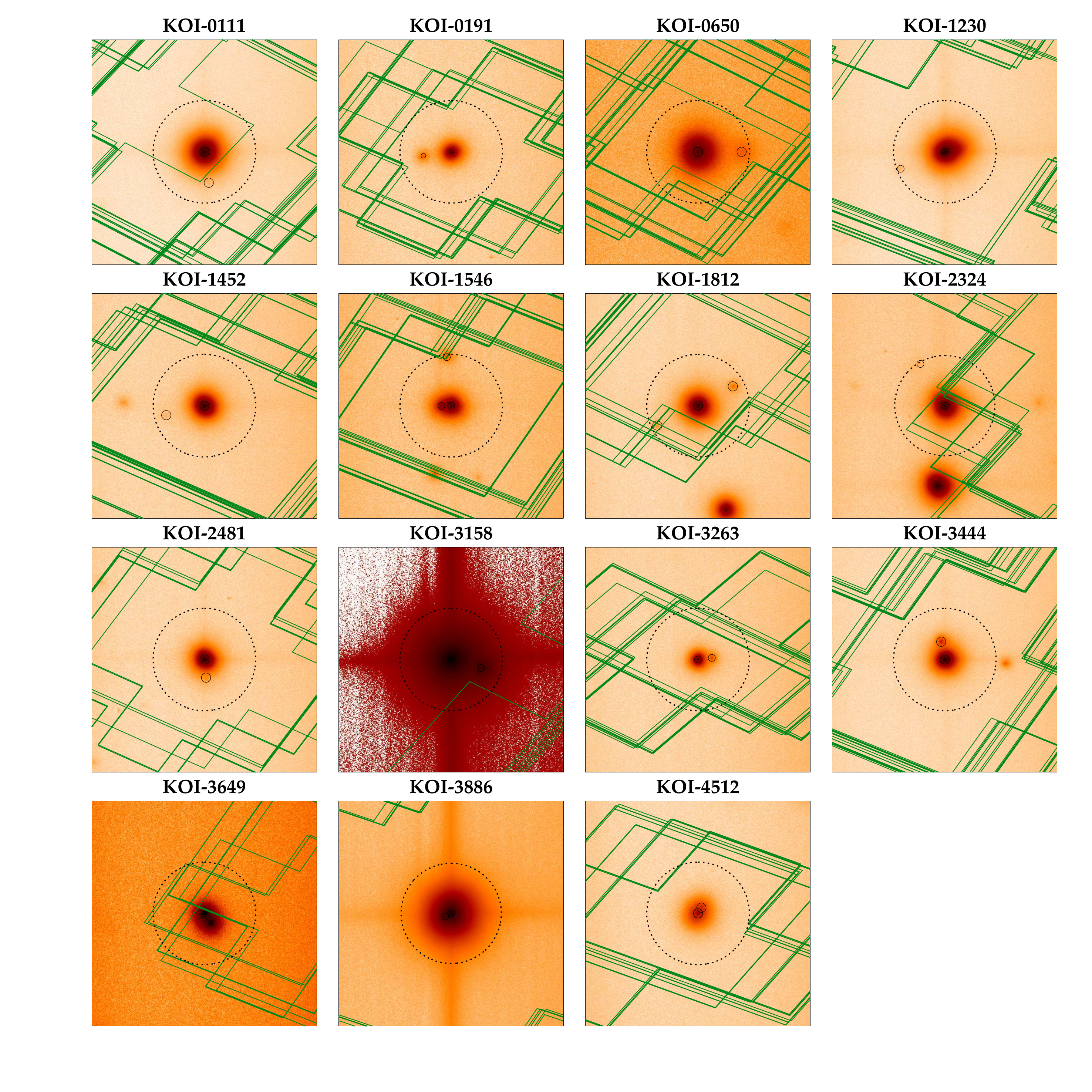}
      \caption{High-resolution images obtained with AstraLux/CAHA of the 15 KOIs with detected companions closer than 3 arcsec. The dotted circles represent the 3 arcsec angular separations and the solid line circles show location of the detected sources. A color version of this figure can be found in the online version of the paper.}
         \label{fig:closecompanions}
   \end{figure*}

\subsubsection{Effect on the planet properties \label{sec:EffectProperties}}

The diluted light from the blended companion that is not accounted in the light curve analysis provides erroneous determinations of the properties of the transiting object. { We note that {\it Kepler} light-curves are deblended from sources detected by either the UKIRT J-band survey or by the KIC photometric survey on the {\it Kepler} field. Among the 15 KOIs with detected companions that are closer than 3 arcsec by the present survey, two of them have all their companions detected by the UKIRT J-band survey (namely KOI-0650 and KOI-1452). Thus, we can remove this KOIs from the current study. In the other two cases (KOI-1546 and KOI-1812), only one out of the two detected companions within 3 arcsec have also been detected by the UKIRT J-band survey. Thus, for these two, we only take the non-detected source into account. Finally, there are another two cases (KOI-2324 and KOI-3886), where the detection of the companions in the UKIRT J-band survey and their correspondence to our detected companions remains unclear. In what follows, we proceed for these targets as if their companions were non-detected by the UKIRT J-band survey. In total, 13 KOIs hosting 24 planet candidates are affected by additional non-accounted fluxes of blended companions.}

In section 4.3 of \cite{lillo-box12} { and section \S~\ref{sec:diluted} of this paper}, we provided theoretical estimations of how the planetary radius of KOIs with detected close companions change because of the detection of blended stars (assuming the brighter object as the host star). In this work, we proceed in the same manner to estimate the corrected planetary radius. The results for the { 24} affected planets in the 2012 and 2013 observed targets are listed in Table~\ref{tab:NewParameters}. The estimations clearly show the increase in the planetary radius caused by the non-accounted flux of the blended star. In { three} cases (namely, KOI-1230.01, KOI-3649.01, and KOI-3886.01), the planet candidate has a new estimated radius that according to mass-radius relations by \cite{chabrier00b} would yield to a typical mass of the transiting object in the stellar regime. We also note that the largest extrasolar planets confirmed so far\footnote{According to http://exoplanet.eu}  have a maximum radius below 2.2 $R_j$. Thus, we can flag these candidates as probable false positives.

We note that other orbital parameters, such as the semi-major axis or the impact parameter, are also affected by the presence of blended stars. However, the correction in these cases as well as a fine correction of the planetary radius involves careful re-analysis of the transit signals, which is out of the scope of this work.


\subsubsection{Physical bond of blended companions \label{sec:zams}}

For those KOIs with observations in both $i$ and $z$ filters, we can further constrain whether the close companions are bound or not to the central object using the color information. We proceed in the same way as in \cite{lillo-box12} (section \S~4.2.2) by testing whether the close companions lie in the same zero-age main-sequence (ZAMS) as the central objects by projecting them to the same distance (see details in the aforementioned paper). Figure~\ref{fig:zams} summarizes the results for the five KOIs with remaining planet candidates and $i,z$ observations during the 2012-2013 observing seasons. Among them, the companion to KOI-3158 (called KOI-3158B) seems to have a consistent age (within the uncertainties) with the main object, since they lie in the same ZAMS. This result agrees with a common formation of the two components, thus being a possible bound binary system. If this were the case, the estimated projected distance between the two components\footnote{Obtained by assuming the distance to the KOI derived when the primary object is located in the empirical ZAMS. We obtain $d=600\pm610$ pc for KOI-3158.} would be $1100\pm1000$ AU for KOI-3158AB. According to our spectral type analysis based on the $i-z$ measured color in section \S~\ref{sec:nonisolated}, the compainon is a redder object in the range K5-M1 (assuming that it is a main-sequence star). 

Although the orbital parameters must be revised to account for the detection of the blended object, we can conclude that KOI-3158 is a  potential candidate to be a S-type planetary systems. Indeed, it has five planet candidates as detected by {\it Kepler}.

   \begin{figure}[ht]
   \centering
   \includegraphics[width=0.5\textwidth]{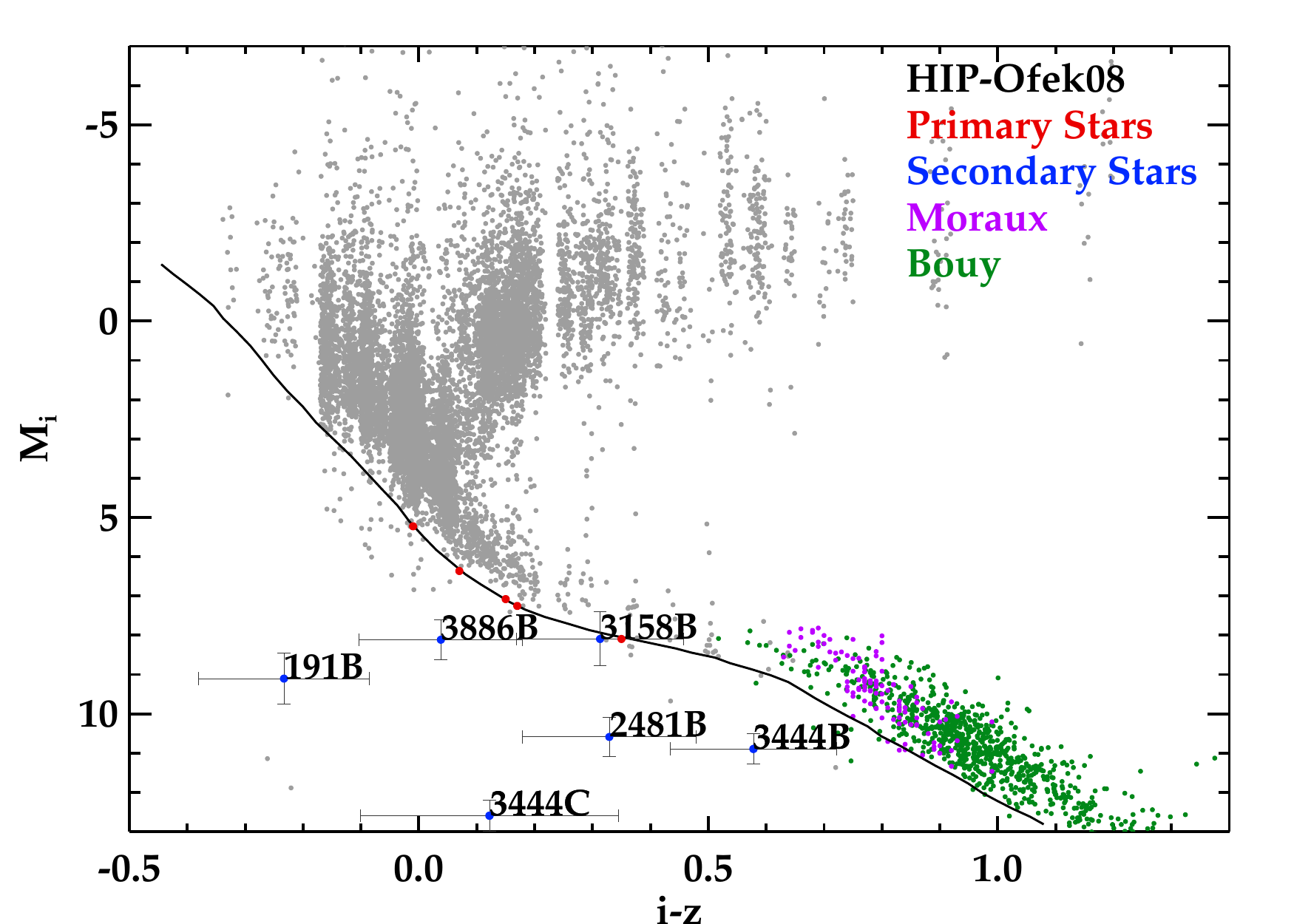}
      \caption{Estimation of the possible bond of the close companions to KOIs detected in both $i$ and $z$ filters during the 2012 and 2013 observing seasons (see section \S~\ref{sec:zams}). The primary sources are represented in red and the close companions in blue. The solid black line represents the empirical ZAMS obtained by using the synthetic $iz$ photometry from \cite{ofek08} as grey dots, the observed members of the Pleiades cluster by \cite{moraux03} as purple circles, and \cite{bouy13} as green circles.}
         \label{fig:zams}
   \end{figure}

According to their position in the Hertzprung-Russell diagram of Fig.~\ref{fig:zams}, the remaining close companions (KOI-0191B, KOI-3886B, KOI-2481B, and KOI-3444C) are probably background sources.

\subsection{A control sample: demoted KOIs}

As was mentioned in section \S~\ref{sec:targets}, 56 KOIs were rejected as candidates after our lucky-imaging observations were performed (and we call them demoted KOIs). We have taken profit of these observations to contrast the multiplicity results. Among the demoted KOIs, we have found that 38 (67.8\%) are actually isolated; 5 (8.9\%) present one source closer than 1.4 arcsec; 11 (19.6\%) present at least one companion closer than 2.5 arcsec; 13 (23.2\%) have at least one source within 3 arcsec; and another 13 KOIs (23.2\%) present at least one object between 3-6 arcsec. These results are summarized in the last section of Table~\ref{tab:results2}. Compared to the real KOIs sample, we can see that these values are similar. Indeed, we found the same rate of isolated targets in both samples. However, there is a slightly higher amount of close companions in this sample compared to the real KOIs. 

Although the detected planets around these  KOIs were rejected, the light curves of those presenting close sources should be re-analyzed by taking this into account. Depending on the causes that led the different works to demote these objects as candidates, some of them could change the properties of the transiting objects and, perhaps, return them back to the planet candidate status.


\section{Comparison to other high spatial resolution techniques in the {\it Kepler} sample \label{sec:comparison}}
 
There are three main techniques that provide high-resolution (diffraction limited) images from the ground: speckle imaging, adaptive optics, and lucky imaging. Regarding the high spatial resolution studies performed for the {\it Kepler} candidates, there are three main works that have provided exhaustive observations of the candidates appart from our survey. \cite{howell11} (hereafter H11) published the first results of the speckle imaging observations for 156 KOIs, using the 3.5m-telescope WIYN on Kitt Peak. \cite{adams12} and \cite{adams13}  (hereafter A12) provided the adaptive optics multiplicity results in the near-infrared regime for a total of 102 KOIs using both the 6.5m Multiple Mirror Telescope (MMT) and the Palomar Hale 5.1m telescope. A posterior, shallow survey by \cite{law13} with Robo-AO (hereafter L13) provided adaptive optics observations for 715 {\it Kepler} candidates, using the robotic Palomar 1.5m telescope \citep{cenko06}.
The results of the current paper added to those of \cite{lillo-box12} complete the set of high spatial resolution techniques by providing lucky imaging for 234 KOIs in the optical range.

The distribution of {\it Kepler} magnitudes is similar for all surveys, peaking L13 and the present work at slightly fainter magnitudes ($m_{kep}\approx 14$) than H11 and A12 ($m_{kep}\approx 12$). In Fig.~\ref{fig:kepmagcomp}, we show this distribution for the four studies.

   \begin{figure}[ht]
   \centering \includegraphics[width=0.5\textwidth]{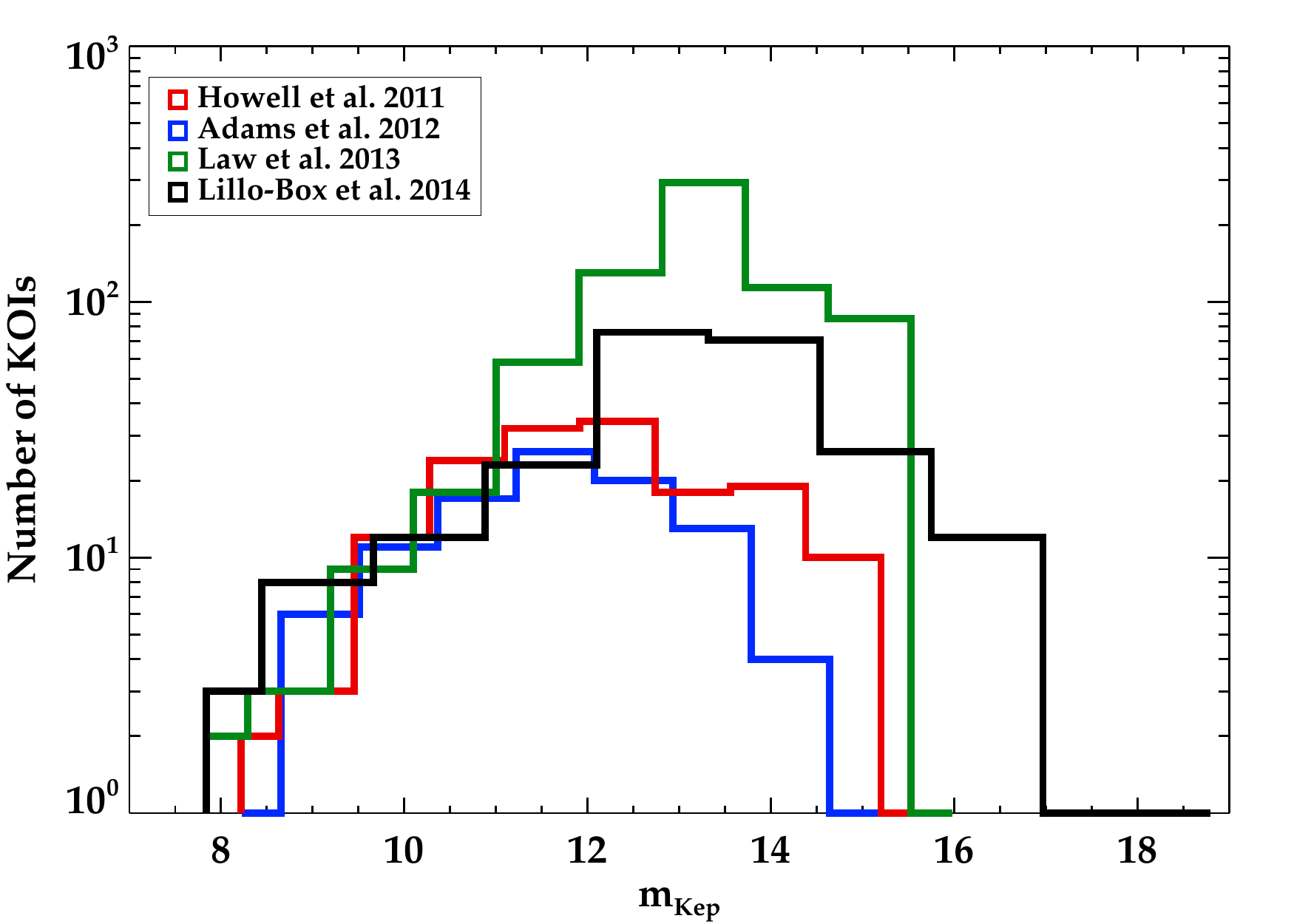}
      \caption{Distributions of {\it Kepler} magnitudes for the different high-spatial resolution surveys in the {\it Kepler} sample of planet host candidates.}
         \label{fig:kepmagcomp}
   \end{figure}

   \begin{figure*}[ht]
   \centering \includegraphics[width=0.32\textwidth]{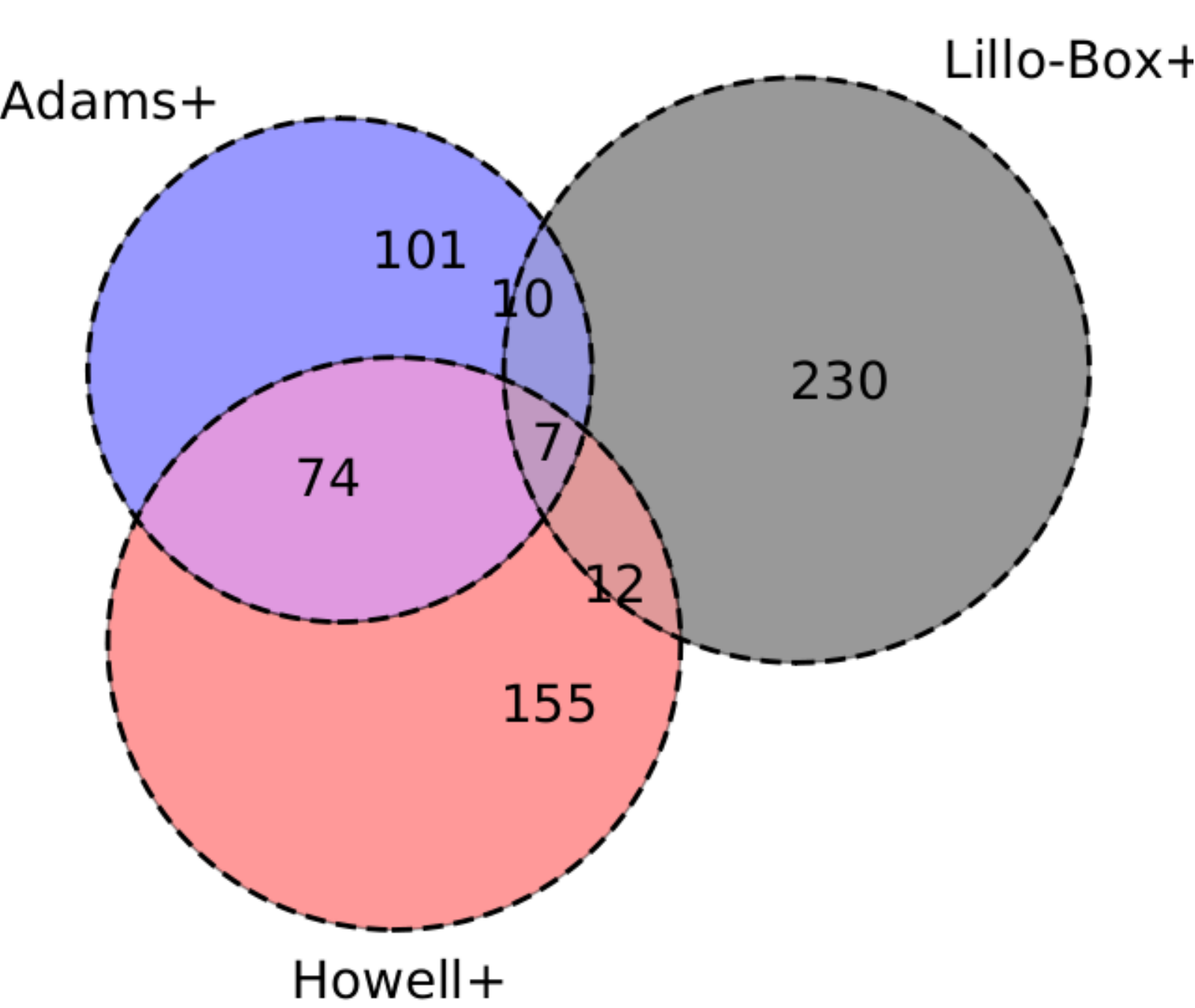}\includegraphics[width=0.32\textwidth]{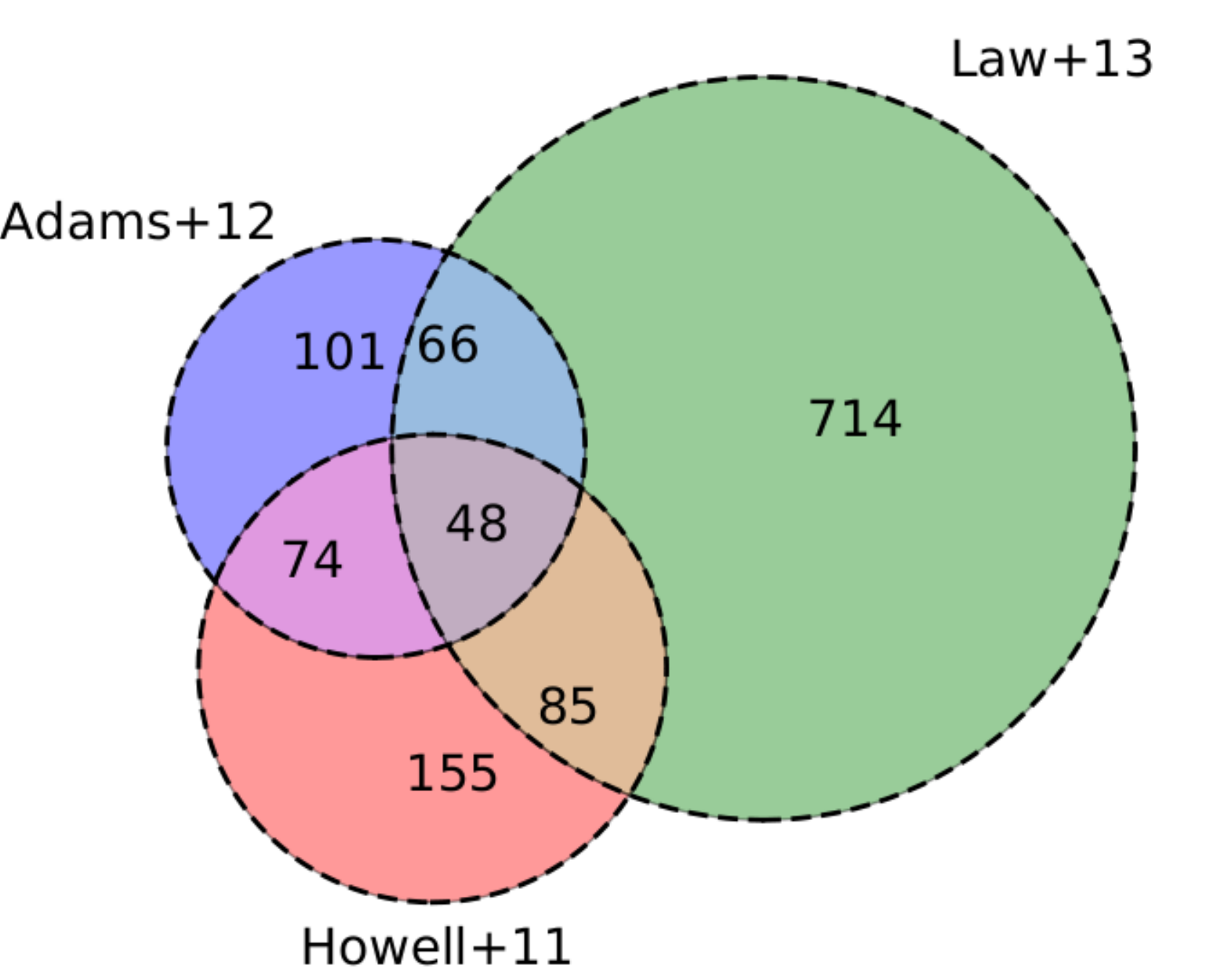}\includegraphics[width=0.32\textwidth]{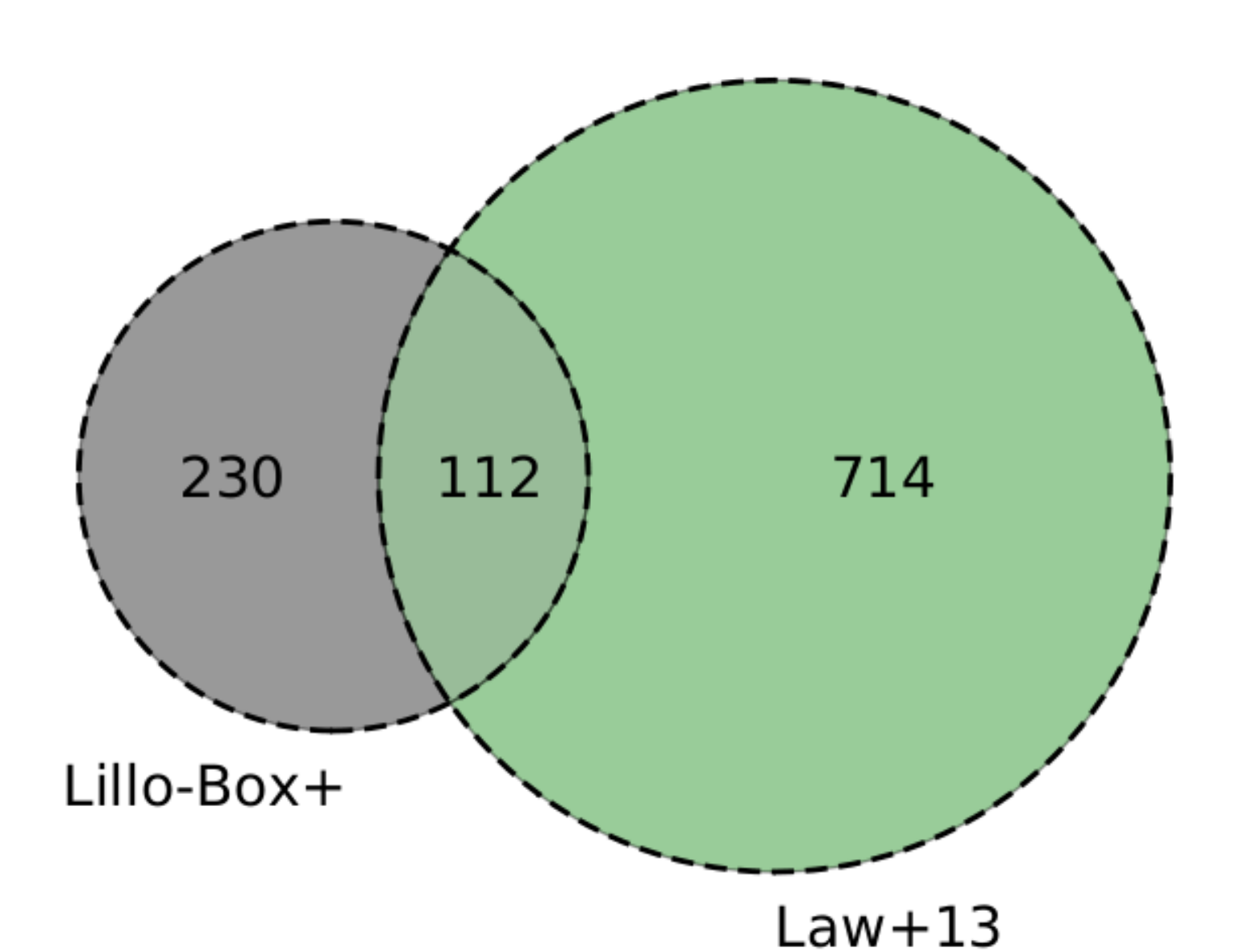}
      \caption{Venn diagrams summarizing the results of the four main high spatial resolution studies regarding the {\it Kepler} sample of planet host candidates.}
         \label{fig:venn}
   \end{figure*}

However, since these works were published, some of their KOIs have been rejected as planet candidates due to several reasons. In particular, 24 KOIs from H11 (out of 156, 15\%), 16 KOIs from A12 (out of 102, 16\%), and 17 KOIs from L13 (out of 714, 2.3\%) currently do not present any planet candidate. In our case, the percentage of non-planet KOIs is slightly higher (56 out of 230, 24\%) because we observed several KOIs, which still had the {\it non-dispositioned} flag in the {\it Kepler} archive (meaning that they did not yet passed all requirements to be planet candidates), in 2013 observing season. 

The speckle imaging study by H11 provides the highest resolution images (with detection limits at 0.05 arcsec) but in a very small field of view that only allows to detect companions at a limiting separation of $1.4$ arcsec. They also provide a typical depth magnitude limitation of $\Delta m=4.0$ mag. The large majority of the transits of planet candidates could be mimicked by blended stellar systems fainter than this magnitude difference (as we can see in the fourth column of Tab.~\ref{tab:bsc}). With these observing limitations, they found that four out of the 127 KOIs (3\%) with remaining planet candidates do present a stellar companion closer than 1.4 arcsec. 

The adaptive optics work by A12 seems to be more complete in magnitude depth and field coverage (more than $20\times 20$ arcsec). We have used their updated Tables 2 and 4 to compute statistics that could be compared to the H11 and our own study. In particular, they find that among their 85 KOIs with remaining planet candidates, 37 are isolated (no companion closer than 6 arcsec, 44 \%), 12 KOIs (14\%) present a stellar companion within 1.4 arcsec, 28 KOIs (33\%) present a stellar companion within 3 arcsec, and 30 KOIs (35\%) present at least one companion in the range 3-6 arcsec. 

The recently published survey by \cite{law13} provides the largest catalog of AO observations of {\it Kepler} candidates. Their observations determine that 29 out of the 697 KOIs with remaining candidates (4.2\%) present some companion within 1.4 arcsec, and 49 (7\%) do present companions closer than 2.5 arcsec. Since this survey is limited to 2.5 arcsec of separation, we cannot include the remaining 648 in the isolated group. 

All these numbers are summarized and compared to the lucky imaging results provided in this paper in Tables~\ref{tab:results2} and \ref{tab:comparisson}.  Figure~\ref{fig:venn} also illustrates the coincident KOIs between the different surveys. In the next subsections, we compare these works to our survey { summarizing coincident objects and BSC results. For the latter, please note that all four studies provide $5\sigma$ level sensitivity limits, so that direct comparisson of the BSC values can be done.}

\subsection{Comparison with \cite{howell11}}

Among the 12 coincident objects between H11 and this work, none  do present companions within 1.4 arcsec (the largest separation detectable by H11). Since sensitivity curves { and photometric transformations of the filters used to the SDSS system} are not provided by the authors for these targets, we cannot exactly compare how both studies have improved the BSC values. As a zero-order approximation, we can assume that the limiting magnitude presented in their Table 2 as $\Delta_{max}$ and calculated for an angular separation of $0.2$ arcsec is constant over the 1.4 arcsec of spatial coverage and obtained in a similar filter\footnote{The filters used by H11 are similar to the I and R Johnson bands.}. We can then determine a zero-order ${\rm BSC}_{H11}$ and thus compare to our values. The results (see Tab.~\ref{tab:h11a12BSC}) show that the speckle-imaging limiting magnitudes are smaller than the required magnitude differences for discarding possible background configurations that are able to mimic the planetary transit (that we have called $\Delta m_{max}$ in this paper). This happens for all 32 planet candidates orbiting the 12 common KOIs. In all cases, our AstraLux observations are better (in terms of reducing the BSC parameter) than the H11 observations.  The small contribution of the H11 study to reduce the probability of a blended eclipsing binary is mostly due to the reduced field of view, which avoids detection of 1.5-3.0 arcsec companions, where the probability of having a background source is maximum in the 0-3 arcsec range.

\subsection{Comparison with \cite{adams12} \label{sec:adams}}

Only four out of the ten coincident objects with \cite{adams12} do present at least one companion below 6 arcsec.

\textit{\textbf{KOI-0111 \& KOI-0555}}. The companions to KOI-0111 and KOI-0555 detected by this work are not detected by A12. Since the magnitude differences in both cases are relatively small ($\Delta_i=6.1$ for KOI-0111 and $\Delta_i=3.8$ for KOI-0555), the non-detection by A12 could mean that these companions are bluer, but we would need photometry in different bands to provide conclusions about this result. 

\textit{\textbf{KOI-0372}}. On the contrary, we do not detect any of the faint companions to KOI-0372 with $\Delta m_{Kep}>10.0$ due to their faintness. However, the maximum magnitude difference for a companion star that could mimic the planetary transit of the candidate KOI-0372.01 is $\Delta m_{max}=4.99$, so that the detected companions by A12 do not affect the planetary candidacy of this object. 

\textit{\textbf{KOI-0115}}. The latter case affects KOI-0115 with three planet candidates for which $\Delta m_{max}(.01)=7.6$, $\Delta m_{max}(.02)=8.8$, and $\Delta m_{max}(.03)=11.1$. The observations from A12 detected two companions with magnitude differences below those values. We do not detect the closest target at 2.43 arcsec and $\Delta m_{kep}=11.4$ mag due to sensitivity restrictions in the present study. However, this companion has a magnitude difference that is higher than the maximum difference that would affect any of the three planets detected in this system. Hence, we could say that this is a negligible blended star for this system.

In the case of the six remaining KOIs with non-detected companions closer than 6 arcsec, we obtain smaller values of the blended source probability, given that our images are deeper for these particular objects. Table~\ref{tab:h11a12BSC} summarizes these results compared to our values according to the updated sensitivity limits that are provided by A12 for each target { in the {\it Kepler} band and transformed to the $i_{SDSS}$ filter using our own transformation determined in section \S~\ref{sec:Ptot}}. In this case, we can see that A12 reduces the probability of having a background source more than H11. The only handicap of this survey is that possible blue non-negligible objects could not be detected by this survey (as we have shown in the cases of KOI-0555 or KOI-0111),  since {\it Kepler} observations are performed in the optical wavelengths and A12 observations are obtained in J and Ks bands.

\subsection{Comparison with \cite{law13}}

A total of 112 KOIs have counterparts in both L13 and the present study. Among this subsample, 13 KOIs have been detected to have companions within 2.5 arcsec (the largest separation that L13 can achieve). In four cases, both studies detect the companions (KOI-0401, KOI-0191, KOI-0628, and KOI-1375). In one case (KOI-0640), our survey does not detect the companion object at 0.44 arcsec with a contrast magnitude of 0.62 mag in the $i$ band. We have examined the AstraLux image and concluded that the ambient conditions were poor for this particular night. This is also reflected in its sensitivity curve with poor quality. Finally, we have detected companions to the remaining eight KOIs (namely, KOI-0658, KOI-1452, KOI-0703, KOI-0704, KOI-0721, KOI-2481, KOI-0111, and KOI-1812) that were not detected by the L13 survey. In L13, the authors justify this non-detection compared to our previous study by arguing that the companions are fainter than their detections limits. However, all planet candidates in these eight planetary systems have calculated a $\Delta m_{max}$ that is fainter than the calculated magnitude differences  of the companion sources. Thus, the detected companions in our survey and those not detected by L13 can actually severely affect the candidacy of the planet candidates or, at least, their planet properties, which are, thus, non-negligible detections. It must be noted that we have detected companions in the range 2.5-6.0 arcsec for another 26 coincident KOIs that could also affect the derived properties of the planet candidate or even their candidacy and that they are not detected in L13 due to field of view restrictions.

Since no individual sensitivity limits are provided for each KOI in L13, we can use their quality definition for each KOI (low, medium, or high), use the correspondent sensitivity curve { in the $i_{SDSS}$ band provided} in their paper to estimate the BSC, and compare it to the values found for our isolated KOIs. The results are presented and compared in Table~\ref{tab:l13BSC}. In general, our observations reduce the $P_{BS}$ by a more significant amount.

\subsection{General comparison of {\it Kepler} high-resolution imaging surveys}

We can compare the results of the surveys by using the BSC parameter defined in previous sections. In particular, we can estimate how each of these high-resolution surveys have contributed to the validation of the planet candidates by measuring how it has diminished the probability of a KOI to have a blended non-detected source. We can calculate the BSC parameter for each observed target in each survey and compare the BSC value prior and after the imaging observations (as we have done in section \S~\ref{sec:Ptot} for the KOIs observed in the present study). We can define the {\it Improvement} parameter as the relative difference between the prior BSC value and the new BSC value obtained with the high-resolution image (i.e. {\it Improvement} $= (P_{BS,0}-P_{BS})/P_{BS,0}$). By doing so, we can summarize the results by the histogram shown in Fig.~\ref{fig:BSCall}. According to this, we can see that A12 obtained a similar distribution of improvements than our work. The only handicap of this survey lies is that the targets were observed in the near-infrared while {\it Kepler} observations are performed in the optical band. Thus, they could miss some bluer companions that affect the {\it Kepler} photometry. This was demonstrated in section \S~\ref{sec:adams} with the cases of KOI-0111 and KOI-0555. {On the contrary, we could be missing redder companions that are possibly bound (such KOI-0372B), which could have implications in the knowledge of the formation and evolution of the planetary system.}

   \begin{figure}[ht]
   \centering 
\includegraphics[width=0.5\textwidth]{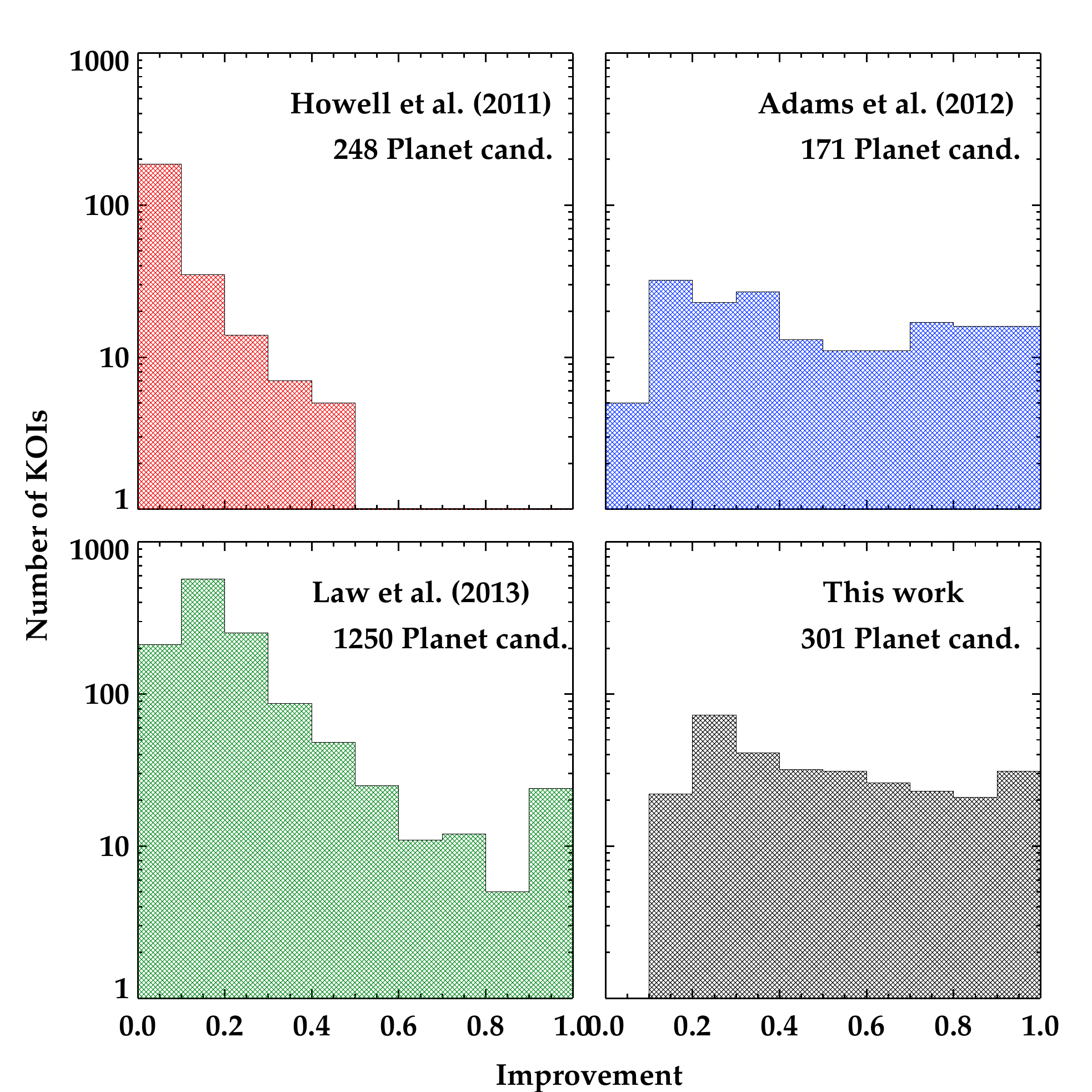}

      \caption{Comparison of the quality of the four main high-resolution surveys of the {\it Kepler} sample of planet candidates. The x-axis represents the improvement in the probability of a background, non-detected companion that could mimic the particular planetary transit.}
         \label{fig:BSCall}
   \end{figure}  

The H11 speckle imaging study does not reduce the probability of a blended scenario in more than 10\% for the large majority of their observations. This is mostly due to the limited contrast magnitude and small field of view that they use. 

In the case of L13, they present the largest sample of high-resolution images, which are also observed in the optical range. Their distribution of improvement for L13 is rather broad. With these observations, the 93\% of the planet candidates hosted by their observed KOIs (1163 planet candidates in total) diminish the probability of a blended scenario by less than 50\%. The remaining 7\% of the planet candidates (87 in total) reduced this probability by more than 50\%. However, since we calculated their $P_{BS}$ by assuming the typical sensitivity curves that are provided by L13 for each target, according to their quality definition of the AO image (namely {\it low}, {\it medium}, and {\it high}), we warn that applying the particular sensitivity curves for each KOI could significantly modify these results.

Finally, our survey provides high-resolution observations for 230 {\it Kepler} host candidates (174 still active KOIs) in the optical range. Our results show improvements in the blended source probability above 50\% for the 62\% of the planet candidates studied (186 in total) and below 50\% for the remaining 38\% of the planet candidates (115 in total).

In Fig.~\ref{fig:comparisson}, we show all the companions detected by the four surveys. Empirical sensitivity curves according to these detections are also plotted for each of the surveys.

   \begin{figure}[ht]
   \centering \includegraphics[width=0.5\textwidth]{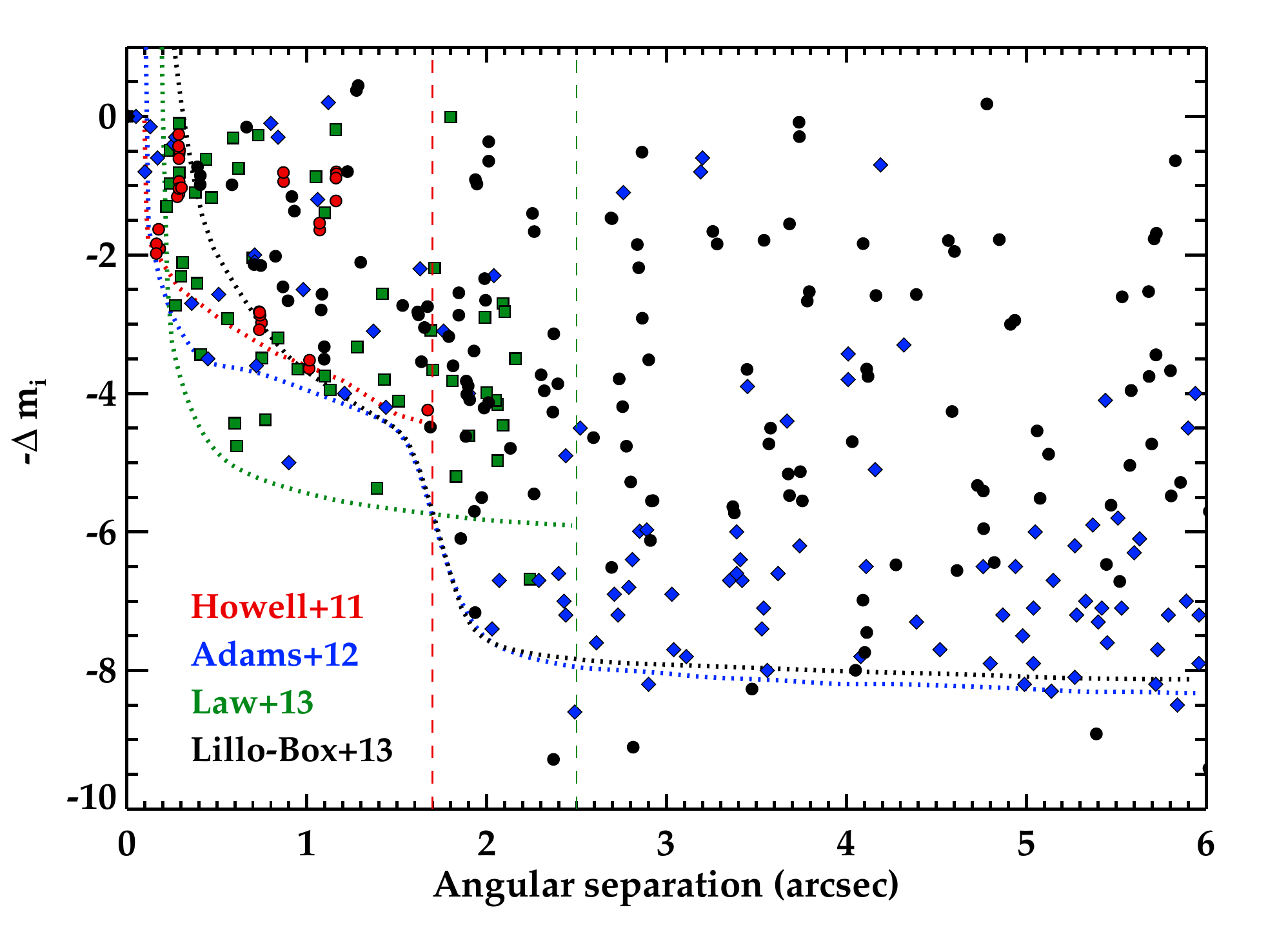}
      \caption{All companions detected by the different high-resolution surveys studied in this work: \cite{adams12} in blue filled diamonds, \cite{howell11} in red filled circles, \cite{law13} in green filled squares, and our work in black filled circles. The empirical sensitivity limits according to these detections are shown with dotted lines.}
         \label{fig:comparisson}
   \end{figure}



\section{Conclusions \label{sec:conclusions} }

In this work, we have delivered a second release of high-resolution observations of {\it Kepler} candidates with the AstraLux/CAHA instrument. In total, 230 KOIs were observed, and 174 currently kept at least one of their planets as candidates. Our complete multiplicity study shows that 111 KOIs (67.2\%) do not present any visual companion closer than 6 arcsec; 35 KOIs (20.1\%) do present at least one source between 3-6 arcsec; and 30 KOIs (17.2\%) present close companions within 3 arcsec. Among the new sample of close companions, we have concluded that KOI-3158B could be physically associated to the planet host, thus being an S-type binary system with multiple planet candidates orbiting one of the components of the binary. These results clearly shows the need for obtaining high-resolution images of planet candidates prior to starting the confirmation process by other (more expensive and time-consuming) techniques, such as radial velocity.

We have analyzed the quality of the images by defining the BSC parameter (background source confidence) that provides the level of confidence by which one can assure that the KOI is isolated within some specific magnitude difference and angular separation. We calculated the BSC parameter for all targets without close companions below 3 arcsec, reducing this probability by more than 50\% for the 62\% of our targets. This implies that thanks to the AstraLux observations we have increased the level of confidence that the KOIs are not blended by an eclipsing binary or affected by additional sources contaminating the {\it Kepler} light curve.

We have performed a comparison to the other three main catalogs of high-resolution images published so far. With the adaptive optics study by \cite{adams12, adams13} our conclusions show that  our work reduces the blended source probability by a high percentage for the majority of the observed targets. 
 The only handicap of the former study is that it is performed in the near-infrared regime; thus, it misses possible hotter companions that could affect the {\it Kepler} photometry obtained in the optical range. This becomes clear in the cases of KOI-0111 and KOI-0555, where we detect close companions non-detected by them. By contrast, we could be missing redder physically bound companions (although they do not affect the planet candidacy since the optical {\it Kepler} light curve would not be significantly affected). The other adaptive optics survey provides a large set of observations performed by a robotic telescope (Robo-AO) by \cite{law13}. They obtained high-resolution images for 714 KOIs in the optical regime. However, although their resolution is high enough, their field of view is too limited (only detecting sources at less than 2.5 arcsec) and their limiting contrast is too small  to improve the isolated confidence by more than 50\% for a significant percentage of their candidates. Hence, this survey must be combined with other observations to achieve useful conclusions. Finally, the speckle study by \cite{howell11} is too limited in both field coverage and magnitude difference, although they achieve very high angular resolutions. This implies that they do not improve the isolation confidence for their targets by more than 10\% and is, thus, not conclusive for this purpose. 

{In this paper, we have also included the high-spatial resolution results for the Kepler-186 (KOI-0571) system, hosting the recently validated planet Kepler-186f, an Earth-like planet in the habitable zone of its cool dwarf \citep{quintana14}. The authors obtained high-contrast images in the Ks band using AO with NIRC2 at the  Keck-II telescope and optical speckle imaging with DSSI at WIYN telescope. The present work adds additional observational support to the conclusion that Kepler-186 is isolated with optical information that is farther than the 1.2 arcsec of the WIYN observations and in the wavelength regime of {\it Kepler} observations.}

Our high-resolution survey of {\it Kepler} candidates with similar important surveys has proven the need for complementary observations of the {\it Kepler} candidates.


\begin{acknowledgements}
      This research has been funded by Spanish grants AYA 2010-21161-C02-02, AYA2012-38897-C02-01, and PRICIT-S2009/ESP-1496. J. Lillo-Box thanks the CSIC JAE-predoc programme for the PhD fellowship. We also thank Calar Alto Observatory, both the open TAC and Spanish GTO panel, for allocating our observing runs, and the CAHA staff for their effort and passion in their work, being very helpful during our visitor and service AstraLux observations. This publication made use of VOSA, developed under the Spanish Virtual Observatory project supported from the Spanish MICINN through grant AyA2008-02156. We also thank the \textit{Kepler} Team for providing valuable information about the targets to be observed. This paper made use of iPython \citep{ipython}, and the python libraries \textit{numpy}, \textit{matplotlib}, \textit{matplotlib\_venn}, and \textit{asciitable}.
\end{acknowledgements}


\bibliographystyle{aa} 
\bibliography{biblio_astralux.bib} 

\begin{thebibliography}{38}
\expandafter\ifx\csname natexlab\endcsname\relax\def\natexlab#1{#1}\fi

\bibitem[{{Adams} {et~al.}(2012){Adams}, {Ciardi}, {Dupree}, {Gautier},
  {Kulesa}, \& {McCarthy}}]{adams12}
{Adams}, E.~R., {Ciardi}, D.~R., {Dupree}, A.~K., {et~al.} 2012, \aj, 144, 42

\bibitem[{{Adams} {et~al.}(2013){Adams}, {Dupree}, {Kulesa}, \&
  {McCarthy}}]{adams13}
{Adams}, E.~R., {Dupree}, A.~K., {Kulesa}, C., \& {McCarthy}, D. 2013, \aj,
  146, 9

\bibitem[{{Barclay} {et~al.}(2013){Barclay}, {Rowe}, {Lissauer}, {Huber},
  {Fressin}, {Howell}, {Bryson}, {Chaplin}, {D{\'e}sert}, {Lopez}, {Marcy},
  {Mullally}, {Ragozzine}, {Torres}, {Adams}, {Agol}, {Barrado}, {Basu},
  {Bedding}, {Buchhave}, {Charbonneau}, {Christiansen},
  {Christensen-Dalsgaard}, {Ciardi}, {Cochran}, {Dupree}, {Elsworth},
  {Everett}, {Fischer}, {Ford}, {Fortney}, {Geary}, {Haas}, {Handberg},
  {Hekker}, {Henze}, {Horch}, {Howard}, {Hunter}, {Isaacson}, {Jenkins},
  {Karoff}, {Kawaler}, {Kjeldsen}, {Klaus}, {Latham}, {Li}, {Lillo-Box},
  {Lund}, {Lundkvist}, {Metcalfe}, {Miglio}, {Morris}, {Quintana}, {Stello},
  {Smith}, {Still}, \& {Thompson}}]{barclay13}
{Barclay}, T., {Rowe}, J.~F., {Lissauer}, J.~J., {et~al.} 2013, \nat, 494, 452

\bibitem[{{Barrado} {et~al.}(2013){Barrado}, {Lillo-Box}, {Bouy}, {Aceituno},
  \& {S{\'a}nchez}}]{barrado13}
{Barrado}, D., {Lillo-Box}, J., {Bouy}, H., {Aceituno}, J., \& {S{\'a}nchez},
  S. 2013, in European Physical Journal Web of Conferences, Vol.~47, European
  Physical Journal Web of Conferences, 5008

\bibitem[{{Batalha} {et~al.}(2010){Batalha}, {Borucki}, {Koch}, {Bryson},
  {Haas}, {Brown}, {Caldwell}, {Hall}, {Gilliland}, {Latham}, {Meibom}, \&
  {Monet}}]{batalha10}
{Batalha}, N.~M., {Borucki}, W.~J., {Koch}, D.~G., {et~al.} 2010, \apjl, 713,
  L109

\bibitem[{{Batalha} {et~al.}(2013){Batalha}, {Rowe}, {Bryson}, {Barclay},
  {Burke}, {Caldwell}, {Christiansen}, {Mullally}, {Thompson}, {Brown},
  {Dupree}, {Fabrycky}, {Ford}, {Fortney}, {Gilliland}, {Isaacson}, {Latham},
  {Marcy}, {Quinn}, {Ragozzine}, {Shporer}, {Borucki}, {Ciardi}, {Gautier},
  {Haas}, {Jenkins}, {Koch}, {Lissauer}, {Rapin}, {Basri}, {Boss}, {Buchhave},
  {Carter}, {Charbonneau}, {Christensen-Dalsgaard}, {Clarke}, {Cochran},
  {Demory}, {Desert}, {Devore}, {Doyle}, {Esquerdo}, {Everett}, {Fressin},
  {Geary}, {Girouard}, {Gould}, {Hall}, {Holman}, {Howard}, {Howell},
  {Ibrahim}, {Kinemuchi}, {Kjeldsen}, {Klaus}, {Li}, {Lucas}, {Meibom},
  {Morris}, {Pr{\v s}a}, {Quintana}, {Sanderfer}, {Sasselov}, {Seader},
  {Smith}, {Steffen}, {Still}, {Stumpe}, {Tarter}, {Tenenbaum}, {Torres},
  {Twicken}, {Uddin}, {Van Cleve}, {Walkowicz}, \& {Welsh}}]{batalha13}
{Batalha}, N.~M., {Rowe}, J.~F., {Bryson}, S.~T., {et~al.} 2013, \apjs, 204, 24

\bibitem[{{Bayo} {et~al.}(2013){Bayo}, {Rodrigo}, {Barrado}, {Solano},
  {Allard}, \& {Joergens}}]{bayo13}
{Bayo}, A., {Rodrigo}, C., {Barrado}, D., {et~al.} 2013, ArXiv e-prints

\bibitem[{{Bayo} {et~al.}(2008){Bayo}, {Rodrigo}, {Barrado Y Navascu{\'e}s},
  {Solano}, {Guti{\'e}rrez}, {Morales-Calder{\'o}n}, \& {Allard}}]{bayo08}
{Bayo}, A., {Rodrigo}, C., {Barrado Y Navascu{\'e}s}, D., {et~al.} 2008, \aap,
  492, 277

\bibitem[{{Bayo} {et~al.}(2014){Bayo}, {Rodrigo}, {Barrado Y Navascu{\'e}s},
  {Solano}, {Guti{\'e}rrez}, {Morales-Calder{\'o}n}, \& {Allard}}]{bayo14}
{Bayo}, A., {Rodrigo}, C., {Barrado Y Navascu{\'e}s}, D., {et~al.} 2014, \aap,
  submitted

\bibitem[{{Borucki} {et~al.}(2011){Borucki}, {Koch}, {Basri}, {Batalha},
  {Brown}, {Bryson}, {Caldwell}, {Christensen-Dalsgaard}, {Cochran}, {DeVore},
  {Dunham}, {Gautier}, {Geary}, {Gilliland}, {Gould}, {Howell}, {Jenkins},
  {Latham}, {Lissauer}, {Marcy}, {Rowe}, {Sasselov}, {Boss}, {Charbonneau},
  {Ciardi}, {Doyle}, {Dupree}, {Ford}, {Fortney}, {Holman}, {Seager},
  {Steffen}, {Tarter}, {Welsh}, {Allen}, {Buchhave}, {Christiansen}, {Clarke},
  {Das}, {D{\'e}sert}, {Endl}, {Fabrycky}, {Fressin}, {Haas}, {Horch},
  {Howard}, {Isaacson}, {Kjeldsen}, {Kolodziejczak}, {Kulesa}, {Li}, {Lucas},
  {Machalek}, {McCarthy}, {MacQueen}, {Meibom}, {Miquel}, {Prsa}, {Quinn},
  {Quintana}, {Ragozzine}, {Sherry}, {Shporer}, {Tenenbaum}, {Torres},
  {Twicken}, {Van Cleve}, {Walkowicz}, {Witteborn}, \& {Still}}]{borucki11}
{Borucki}, W.~J., {Koch}, D.~G., {Basri}, G., {et~al.} 2011, \apj, 736, 19

\bibitem[{{Bouy} {et~al.}(2013){Bouy}, {Bertin}, {Moraux}, {Cuillandre},
  {Bouvier}, {Barrado}, {Solano}, \& {Bayo}}]{bouy13}
{Bouy}, H., {Bertin}, E., {Moraux}, E., {et~al.} 2013, \aap, 554, A101

\bibitem[{{Brown} {et~al.}(2011){Brown}, {Latham}, {Everett}, \&
  {Esquerdo}}]{brown11}
{Brown}, T.~M., {Latham}, D.~W., {Everett}, M.~E., \& {Esquerdo}, G.~A. 2011,
  \aj, 142, 112

\bibitem[{{Burke} {et~al.}(2013){Burke}, {Bryson}, {Christiansen}, {Mullally},
  {Rowe}, {Science Office}, \& {Kepler Science Team}}]{burke13}
{Burke}, C.~J., {Bryson}, S., {Christiansen}, J., {et~al.} 2013, in American
  Astronomical Society Meeting Abstracts, Vol. 221, American Astronomical
  Society Meeting Abstracts, 216.02

\bibitem[{{Cabrera} {et~al.}(2014){Cabrera}, {Csizmadia}, {Lehmann}, {Dvorak},
  {Gandolfi}, {Rauer}, {Erikson}, {Dreyer}, {Eigm{\"u}ller}, \&
  {Hatzes}}]{cabrera13}
{Cabrera}, J., {Csizmadia}, S., {Lehmann}, H., {et~al.} 2014, \apj, 781, 18

\bibitem[{{Cenko} {et~al.}(2006){Cenko}, {Fox}, {Moon}, {Harrison}, {Kulkarni},
  {Henning}, {Guzman}, {Bonati}, {Smith}, {Thicksten}, {Doyle}, {Petrie},
  {Gal-Yam}, {Soderberg}, {Anagnostou}, \& {Laity}}]{cenko06}
{Cenko}, S.~B., {Fox}, D.~B., {Moon}, D.-S., {et~al.} 2006, \pasp, 118, 1396

\bibitem[{{Chabrier}(2001)}]{chabrier01}
{Chabrier}, G. 2001, \apj, 554, 1274

\bibitem[{{Chabrier} \& {Baraffe}(2000)}]{chabrier00b}
{Chabrier}, G. \& {Baraffe}, I. 2000, \araa, 38, 337

\bibitem[{{Daemgen} {et~al.}(2009){Daemgen}, {Hormuth}, {Brandner}, {Bergfors},
  {Janson}, {Hippler}, \& {Henning}}]{daemgen09}
{Daemgen}, S., {Hormuth}, F., {Brandner}, W., {et~al.} 2009, \aap, 498, 567

\bibitem[{{Hormuth}(2007)}]{hormuth07}
{Hormuth}, F. 2007, AstraLux diploma thesis (University of Heidelberg)

\bibitem[{{Howell} {et~al.}(2011){Howell}, {Everett}, {Sherry}, {Horch}, \&
  {Ciardi}}]{howell11}
{Howell}, S.~B., {Everett}, M.~E., {Sherry}, W., {Horch}, E., \& {Ciardi},
  D.~R. 2011, \aj, 142, 19

\bibitem[{{Kley}(2010)}]{kley10}
{Kley}, W. 2010, in EAS Publications Series, Vol.~42, EAS Publications Series,
  ed. {K.~Go{\.z}dziewski, A.~Niedzielski, \& J.~Schneider}, 227--238

\bibitem[{{Law} {et~al.}(2013){Law}, {Morton}, {Baranec}, {Riddle},
  {Ravichandran}, {Ziegler}, {Johnson}, {Tendulkar}, {Bui}, {Burse}, {Das},
  {Dekany}, {Kulkarni}, {Punnadi}, \& {Ramaprakash}}]{law13}
{Law}, N.~M., {Morton}, T., {Baranec}, C., {et~al.} 2013, ArXiv e-prints

\bibitem[{{Lillo-Box} {et~al.}(2012){Lillo-Box}, {Barrado}, \&
  {Bouy}}]{lillo-box12}
{Lillo-Box}, J., {Barrado}, D., \& {Bouy}, H. 2012, \aap, 546, A10

\bibitem[{{Lillo-Box} {et~al.}(2014){Lillo-Box}, {Barrado}, {Moya},
  {Montesinos}, {Montalb{\'a}n}, {Bayo}, {Barbieri}, {R{\'e}gulo}, {Mancini},
  {Bouy}, \& {Henning}}]{lillo-box13}
{Lillo-Box}, J., {Barrado}, D., {Moya}, A., {et~al.} 2014, \aap, 562, A109

\bibitem[{{Lissauer} {et~al.}(2014){Lissauer}, {Marcy}, {Bryson}, {Rowe},
  {Jontof-Hutter}, {Agol}, {Borucki}, {Carter}, {Ford}, {Gilliland}, {Kolbl},
  {Star}, {Steffen}, \& {Torres}}]{lissauer14}
{Lissauer}, J.~J., {Marcy}, G.~W., {Bryson}, S.~T., {et~al.} 2014, \apj, 784,
  44

\bibitem[{{Marcy} {et~al.}(2014){Marcy}, {Isaacson}, {Howard}, {Rowe},
  {Jenkins}, {Bryson}, {Latham}, {Howell}, {Gautier}, {Batalha}, {Rogers},
  {Ciardi}, {Fischer}, {Gilliland}, {Kjeldsen}, {Christensen-Dalsgaard},
  {Huber}, {Chaplin}, {Basu}, {Buchhave}, {Quinn}, {Borucki}, {Koch}, {Hunter},
  {Caldwell}, {Van Cleve}, {Kolbl}, {Weiss}, {Petigura}, {Seager}, {Morton},
  {Johnson}, {Ballard}, {Burke}, {Cochran}, {Endl}, {MacQueen}, {Everett},
  {Lissauer}, {Ford}, {Torres}, {Fressin}, {Brown}, {Steffen}, {Charbonneau},
  {Basri}, {Sasselov}, {Winn}, {Sanchis-Ojeda}, {Christiansen}, {Adams},
  {Henze}, {Dupree}, {Fabrycky}, {Fortney}, {Tarter}, {Holman}, {Tenenbaum},
  {Shporer}, {Lucas}, {Welsh}, {Orosz}, {Bedding}, {Campante}, {Davies},
  {Elsworth}, {Handberg}, {Hekker}, {Karoff}, {Kawaler}, {Lund}, {Lundkvist},
  {Metcalfe}, {Miglio}, {Silva Aguirre}, {Stello}, {White}, {Boss}, {Devore},
  {Gould}, {Prsa}, {Agol}, {Barclay}, {Coughlin}, {Brugamyer}, {Mullally},
  {Quintana}, {Still}, {Thompson}, {Morrison}, {Twicken}, {D{\'e}sert},
  {Carter}, {Crepp}, {H{\'e}brard}, {Santerne}, {Moutou}, {Sobeck}, {Hudgins},
  {Haas}, {Robertson}, {Lillo-Box}, \& {Barrado}}]{marcy14}
{Marcy}, G.~W., {Isaacson}, H., {Howard}, A.~W., {et~al.} 2014, \apjs, 210, 20

\bibitem[{{Moraux} {et~al.}(2003){Moraux}, {Bouvier}, {Stauffer}, \&
  {Cuillandre}}]{moraux03}
{Moraux}, E., {Bouvier}, J., {Stauffer}, J.~R., \& {Cuillandre}, J.-C. 2003,
  \aap, 400, 891

\bibitem[{{Morton} \& {Johnson}(2011)}]{morton11}
{Morton}, T.~D. \& {Johnson}, J.~A. 2011, \apj, 738, 170

\bibitem[{{Ofek}(2008)}]{ofek08}
{Ofek}, E.~O. 2008, \pasp, 120, 1128

\bibitem[{P\'erez \& Granger(2007)}]{ipython}
P\'erez, F. \& Granger, B.~E. 2007, {C}omput. {S}ci. {E}ng., 9, 21

\bibitem[{{Pinsonneault} {et~al.}(2012){Pinsonneault}, {An},
  {Molenda-{\.Z}akowicz}, {Chaplin}, {Metcalfe}, \& {Bruntt}}]{pinsonneault12}
{Pinsonneault}, M.~H., {An}, D., {Molenda-{\.Z}akowicz}, J., {et~al.} 2012,
  \apjs, 199, 30

\bibitem[{Quintana {et~al.}(2014)Quintana, Barclay, Raymond, Rowe, Bolmont,
  Caldwell, Howell, Kane, Huber, Crepp, Lissauer, Ciardi, Coughlin, Everett,
  Henze, Horch, Isaacson, Ford, Adams, Still, Hunter, Quarles, \&
  Selsis}]{quintana14}
Quintana, E.~V., Barclay, T., Raymond, S.~N., {et~al.} 2014, Science, 344, 277

\bibitem[{{Quintana} {et~al.}(2013){Quintana}, {Rowe}, {Barclay}, {Howell},
  {Ciardi}, {Demory}, {Caldwell}, {Borucki}, {Christiansen}, {Jenkins},
  {Klaus}, {Fulton}, {Morris}, {Sanderfer}, {Shporer}, {Smith}, {Still}, \&
  {Thompson}}]{quintana13}
{Quintana}, E.~V., {Rowe}, J.~F., {Barclay}, T., {et~al.} 2013, \apj, 767, 137

\bibitem[{{Rowe} {et~al.}(2014){Rowe}, {Bryson}, {Marcy}, {Lissauer},
  {Jontof-Hutter}, {Mullally}, {Gilliland}, {Issacson}, {Ford}, {Howell},
  {Borucki}, {Haas}, {Huber}, {Steffen}, {Thompson}, {Quintana}, {Barclay},
  {Still}, {Fortney}, {Gautier}, {Hunter}, {Caldwell}, {Ciardi}, {Devore},
  {Cochran}, {Jenkins}, {Agol}, {Carter}, \& {Geary}}]{rowe14}
{Rowe}, J.~F., {Bryson}, S.~T., {Marcy}, G.~W., {et~al.} 2014, \apj, 784, 45

\bibitem[{{Steffen} {et~al.}(2012){Steffen}, {Ford}, {Rowe}, {Fabrycky},
  {Holman}, {Welsh}, {Batalha}, {Borucki}, {Bryson}, {Caldwell}, {Ciardi},
  {Jenkins}, {Kjeldsen}, {Koch}, {Pr{\v s}a}, {Sanderfer}, {Seader}, \&
  {Twicken}}]{steffen12}
{Steffen}, J.~H., {Ford}, E.~B., {Rowe}, J.~F., {et~al.} 2012, \apj, 756, 186

\bibitem[{{Strehl}(1902)}]{strehl1902}
{Strehl}, K. 1902, Astronomische Nachrichten, 158, 89

\bibitem[{{Weiss} {et~al.}(2013){Weiss}, {Marcy}, {Rowe}, {Howard}, {Isaacson},
  {Fortney}, {Miller}, {Demory}, {Fischer}, {Adams}, {Dupree}, {Howell},
  {Kolbl}, {Johnson}, {Horch}, {Everett}, {Fabrycky}, \& {Seager}}]{weiss13}
{Weiss}, L.~M., {Marcy}, G.~W., {Rowe}, J.~F., {et~al.} 2013, \apj, 768, 14

\bibitem[{{Yanny} {et~al.}(1994){Yanny}, {Guhathakurta}, {Bahcall}, \&
  {Schneider}}]{yanny94}
{Yanny}, B., {Guhathakurta}, P., {Bahcall}, J.~N., \& {Schneider}, D.~P. 1994,
  \aj, 107, 1745

\end{thebibliography}

\clearpage



\longtab{1}{
\scriptsize 
\begin{longtable}{crccccccccccc}
\caption{Observational information, completeness, and detectability magnitudes for the isolated KOIs observed with the Calar Alto/AstraLux instrument during the 2011 observational season (62 KOIs in total) and all targets observed during 2012 and 2013 seassons (20 and 57 respectively KOIs). Note that the targets with close companions from 2011 (another 41 KOIs)  were already published in \cite{lillo-box12}. Active KOIs (currently keeping at least one of their planet candidates) are shown in the upper part of the table and demoted KOIs are shown in the bottom part of the table.}\\             
\label{tab:observations} \\     
\hline\hline
\multicolumn{13}{c}{\textbf{Active KOIs}} \\ \hline
KOI & Others\tablefootmark{a} & KIC & RA\tablefootmark{b} & DEC\tablefootmark{b} & kep & Date & Filter &$T_{ind}$ & \#Frames & Eff.Time & $i_{comp}$\tablefootmark{c}& $i_{det}$\tablefootmark{d}   \\
ID & & & J2000.0 & J2000.0 & mag & yyyy-mm-dd & &   \centering{$s$} & & s & mag. & mag. \\
\hline
\endfirsthead
\caption{continued.}\\
\hline\hline
KOI & Others\tablefootmark{a} & KIC & RA\tablefootmark{b} & DEC\tablefootmark{b} & kep & Date & Filter &$T_{ind}$ & \#Frames & Eff.Time & $i_{comp}$\tablefootmark{c}& $i_{det}$\tablefootmark{d} \\
ID & & & J2000.0 & J2000.0 & mag & yyyy-mm-dd & &   \centering{$s$} & & s & mag. & mag.   \\
\hline
\endhead
\hline
\endfoot

  12 &     L  & 5812701 & 19:49:48.9 & 41:00:39.56 & 11.353 & 2011-06-03 &  i & 0.05 & 10000 & 50.0 & 16.9 & 20.2\\
  41 & H,A,L  & 6521045 & 19:25:32.64 & 41:59:24.97 & 11.197 & 2013-06-22 &  i & 0.083 & 40000 & 332.0 & 19.0 & 22.3\\
  49 &   H,L  & 9527334 & 19:28:59.77 & 46:09:53.36 & 13.704 & 2013-06-22 &  i & 0.083 & 40000 & 332.0 & 19.0 & 22.3\\
  51 &     -  & 6056992 & 19:43:40.52 & 41:19:56.76 & 13.761 & 2013-06-22 &  i & 0.06 & 20000 & 120.0 & 17.8 & 21.1\\
  69 & H,A,L  & 3544595 & 19:25:40.39 & 38:40:20.49 & 9.931 & 2013-06-22 &  i & 0.045 & 40000 & 180.0 & 18.3 & 22.6\\
  82 & H,A,L  & 10187017 & 18:45:55.85 & 47:12:28.91 & 11.492 & 2013-06-23 &  i & 0.068 & 40000 & 272.0 & 18.7 & 22.0\\
  94 & H,A,L  & 6462863 & 19:49:19.94 & 41:53:28.04 & 12.205 & 2013-06-21 &  i & 0.06 & 30000 & 180.0 & 18.3 & 21.6\\
 111 & H,A,L  & 6678383 & 19:10:25.11 & 42:10:00.4 & 12.596 & 2013-06-23 &  i & 0.068 & 40000 & 272.0 & 18.7 & 22.0\\
\end{longtable}
\tablefoot{ \\
\tablefoottext{a}{Identifier of papers where high-resolution images are provided for each KOI: H for \cite{howell11}, A for \cite{adams12}, and L for \cite{law13}.}\\
\tablefoottext{b}{Right ascension and declination from \cite{borucki11}.}\\
\tablefoottext{c}{Estimated completeness magnitudes scaled to those found for the 200 s image of the globular cluster M15.}\\
\tablefoottext{d}{Estimated detectability magnitudes scaled to those found for the 200 s image of the globular cluster M15.}
}

}

\begin{table*}
\caption{Plate solution for our photometric observations (see section \S~\ref{sec:astrometry}).}             
\scriptsize
\setlength{\extrarowheight}{3pt}
\label{tab:astrometry}      
\centering                          
\begin{tabular}{c c| c |  c|  c }        
\hline\hline                 

 Parameter      & Units  & 2011 &  2012 &  2013  \\
\hline                        
\# stars &           &   100   &   179  &    239  \\ 
Pixel scale & $mas/px$ &  $23.89\pm0.23$     &   $23.89\pm0.23$  &    $23.61\pm0.20$   \\
PA      & degrees    &  $1.74\pm0.54$   &     $1.94\pm0.57$  & $1.96\pm0.50$    \\


\hline                                   
\end{tabular}
\end{table*}

\longtab{3}{
\scriptsize 
\begin{longtable}{r c c c c c c c c c c r}
\caption{Photometry of the 23 active KOIs (upper part of the table) and 18 demoted KOIs (bottom part of the table) with detected companions that are closer than 6 arcsec in the new sample observed during 2012 and 2013.}             
\label{tab:photometry}  \\    
\hline\hline
\multicolumn{12}{c}{\textbf{Active KOIs}} \\ \hline 
KOI & Comp. &   Ang.Sep. &      Angle &                SDSSi &                SDSSz &           $\Delta i$ &           $\Delta z$ &                  i-z &                  SpT &    Season & Others\tablefootmark{a} \\
    &       &     arcsec &       deg. &                 mag. &                 mag. &                 mag. &                 mag. &                 mag. &                      &           &           \\        
\hline
\endfirsthead
\caption{continued.}\\
\hline\hline
KOI & Comp. &   Ang.Sep. &      Angle &                SDSSi &                SDSSz &           $\Delta i$ &           $\Delta z$ &                  i-z &                  SpT &    Season & Others\tablefootmark{a} \\
    &       &     arcsec &       deg. &                 mag. &                 mag. &                 mag. &                 mag. &                 mag. &                      &           &           \\       
\hline
\endhead
\hline
\endfoot

       111 &     A &        $0.0$         &        $0.0$         &       $12.38\pm0.10$ &              $-$     &      $0.0$           &              $-$     &              $-$     &       $G0^{G5}_{F8}$ &       2013 &   H,A,L  \\
       111 &     B &      $1.856\pm0.018$ &      $186.11\pm0.50$ &       $18.47\pm0.54$ &              $-$     &     $-6.095\pm0.526$ &              $-$     &              $-$     &                    - &       2013 &   L      \\
       115 &     A &        $0.0$         &        $0.0$         &       $12.59\pm0.10$ &              $-$     &      $0.0$           &              $-$     &              $-$     &       $F8^{G2}_{F5}$ &       2013 &   A,L    \\
       115 &     B &      $4.051\pm0.039$ &      $337.43\pm0.50$ &       $20.59\pm1.50$ &              $-$     &     $-7.997\pm2.251$ &              $-$     &              $-$     &                    - &       2013 &   A,L    \\
       152 &     A &        $0.0$         &        $0.0$         &       $13.69\pm0.10$ &              $-$     &      $0.0$           &              $-$     &              $-$     &       $G0^{G5}_{F8}$ &       2013 &   L      \\
       152 &     B &      $5.721\pm0.056$ &      $340.17\pm0.50$ &       $17.14\pm0.23$ &              $-$     &     $-3.443\pm0.044$ &              $-$     &              $-$     &                    - &       2013 &   -      \\
\end{longtable}
\tablefoot{\\
\tablefoottext{a}{Identifier of papers that have detected the KOI and/or the companion: H for \cite{howell11}, A for \cite{adams12}, and L for \cite{law13}.}
}}

\longtab{4}{
\scriptsize 
\begin{longtable}{rrccccccc}
\caption{Sensitivity limit results { (at 5$\sigma$ level)} for all observed targets (230 KOIs) in the 2011, 2012, and 2013 observing seasons. {The sensitivity limits at the different angular separations ($\Delta m$, Columns 5 to 9) are provided in the corresponding $i$ and $z$ bands (Column 3).}} \\             
\label{tab:sensitivity} \\     
\hline\hline
 \multicolumn{8}{c}{\textbf{Active KOIs}} \\ \hline 
      KOI  &   Others\tablefootmark{a} &   Filter  & m\tablefootmark{b}  &   $\Delta m$ @ 0.2''  &     $\Delta m$ @ 0.5''  &     $\Delta m$ @ 1.0''  &     $\Delta m$ @ 2.0''  &     $\Delta m$ @ 3.0''  \\
\hline
\endfirsthead
\caption{continued.}\\
\hline\hline
       KOI  &  Others\tablefootmark{a} &   Filter  & m\tablefootmark{b}  &   $\Delta m$ @ 0.2''  &     $\Delta m$ @ 0.5''  &     $\Delta m$ @ 1.0''  &     $\Delta m$ @ 2.0''  &     $\Delta m$ @ 3.0''  \\
\hline
\endhead
\hline
\endfoot

        12  &       L     &    i       &                  11.19   &   1.99   &   3.58  &    4.20  &    3.40  &    4.50  \\
        41  &   H,A,L     &    i       &                  10.98   &   0.08   &   2.32  &    5.64  &    8.55  &    8.71	\\
        49  &     H,L     &    i       &                  13.46   &   0.75   &   3.61  &    6.78  &    8.11  &    6.83	\\
        51  &       -     &    i       &                  13.98   &   1.11   &   3.22  &    5.93  &    7.64  &    6.50	\\
        69  &   H,A,L     &    i       &                   9.69   &   2.50   &   1.46  &    4.87  &    7.82  &    9.00	\\
        82  &   H,A,L     &    i       &                  11.11   &   0.87   &   2.74  &    5.28  &    8.17  &    8.83	\\
        94  &   H,A,L     &    i       &                  12.01   &   1.52   &   2.88  &    4.64  &    7.05  &    7.72	\\
        99  &       L     &    i       &                  12.64   &   1.08   &   3.91  &    6.96  &    7.57  &    6.20	\\
\end{longtable}
\tablefoot{\\
\tablefoottext{a}{Identifier of papers where high-resolution images are provided for each KOI: H for \cite{howell11}, A for \cite{adams12}, and L for \cite{law13}.}\\
\tablefoottext{b}{Magnitude in the SDSS filter system of the corresponding band (in previous column) obtained by transforming the KIC magnitudes with equations provided by \cite{pinsonneault12}. }\\
\textdagger KIC magnitude that was not converted to the SDSS system due to the lack of some needed magnitudes in equations 3 and 4 of \cite{pinsonneault12}.
}
}

\begin{table*}
\caption{Multiplicity results. The table shows the number of KOIs with at least one companion closer than 3 arcsec and 3-6 arcsec and those without companions closer than 6 arcsec within our observational limitations (see section \S~\ref{sec:sensitivity}) for each season. The lower separation limit of 0.1 arcsec is due to the minimum achievable resolution with AstraLux in optimum weather conditions.  }             
\setlength{\extrarowheight}{7pt}
\label{tab:results}      
\centering                          
\begin{tabular}{c c c c c}        
\hline\hline                 

Run            &  Observed & Isolated   &  3-6 arcsec &  0.1-3 arcsec \\ \hline
2011 (update)  &   97      &    63 (65.0\%) & 22 (22.7\%)  & 15 (15.8\%) \\
2012+2013      &   77      &    54 (68.4\%) & 12 (15.6\%)  & 15 (19.7\%) \\ \hline
{\bf Total}          &  {\bf 174}      &   {\bf 117 (67.2\%)} & {\bf 34 (19.5\%)}  & {\bf 30 (17.2\%)} \\

\hline                                   
\end{tabular}
\end{table*}

\begin{table*}
\caption{Multiplicity results for the four main works on high-resolution imaging on the Kepler sample of candidates. We show the number of detected companions for different separation ranges. The lower part of the table shows the statistics regardless whether the KOI still hosts planet candidates. We have 56 KOIs that have been demoted in the latest {\it Kepler} releases and are not classified as planet hosts any longer. In the case of \cite{howell11}, there are 24 demoted KOIs, there are 16 for \cite{adams12,adams13}, and there are 17 for \cite{law13}. These results are presented int the third section of this table.}             
\setlength{\extrarowheight}{7pt}
\label{tab:results2}      
\centering                          
\begin{tabular}{l c c r c c c c}        
\hline\hline                 

\multicolumn{8}{c}{\textbf{Targets with remaining planet candidates (current valid KOIs)}} \\ \hline
Study                       & Technique\tablefootmark{a}     &  Observed & Isolated   &  0.0\arcsec-1.4\arcsec & 0.0\arcsec-2.5\arcsec  & 0.0\arcsec-3.0\arcsec &  3\arcsec-6\arcsec \\ \hline
Howell et al. 2011          & speckle, opt                   &   131     &    ---         &  4  (3\%)      & ---         & ---         &  ---         \\   
Adams et al. 2012,2013      & AO, near-IR                    &   85      &    37 (44\%)   &  12 (14\%)     & 23 (27\%)   & 28 (33\%)   &  30 (35\%)   \\ 
Law et al. 2013             & AO, opt                        &   697     &    ---         &  29 (4\%)      & 49 (7\%)    & ---         & ---          \\
Lillo-Box et al. 2011-2013  & lucky, opt                     &   174     &    117 (67\%)  &  9  (5\%)      & 25 (14\%)   & 30 (17\%)   &  34 (20\%)   \\ \hline \hline
\multicolumn{8}{c}{\textbf{All targets observed (valid and demoted KOIs)}} \\ \hline
Study                       & Technique\tablefootmark{a}     &  Observed & Isolated   &  0.0\arcsec-1.4\arcsec & 0.0\arcsec-2.5\arcsec  & 0.0\arcsec-3.0\arcsec &  3\arcsec-6\arcsec \\
\end{tabular}
\tablefoot{\\
\tablefoottext{a}{Technique and wavelength range (opt = optical, near-IR = near-infrared) used in the study.}\\
}
\end{table*}

\longtab{7}{
\scriptsize 
\begin{longtable}{rcccccccc}
\caption{Results of the blended source probability prior ($P_{BS,0}$) and after ($P_{BS}^{LB14}$) the AstraLux images for 222 planet candidates around isolated KOIs from all three observing runs ($Type=0$) and KOIs with detected companions at 3-6 arcsec from the 2012 and 2013 observing runs ($Type=1$). See section \S~\ref{sec:bsc} for more details.} \\             
\label{tab:bsc} \\     
\hline\hline
\multicolumn{9}{c}{\textbf{Active KOIs}} \\ \hline 
         Planet &  Type\tablefootmark{a} &  $m_{kep}$ &  $\Delta m_{max}$ &        $P_{BS,0}$ &  $P_{BS}^{LB14}$ &         Improve &       $P_{appEB}$ &          $P_{BB}$  \\
      candidate &       &        mag &               mag &              \% &                   \% &              \% &              \% &              \%  \\
\hline
\endfirsthead
\caption{continued.}\\
\hline\hline
         Planet &  Type\tablefootmark{a} &  $m_{kep}$ &  $\Delta m_{max}$ &        $P_{BS,0}$ &  $P_{BS}^{LB14}$ &         Improve &       $P_{appEB}$ &          $P_{BB}$  \\
      candidate &       &        mag &               mag &              \% &                   \% &              \% &              \% &              \%  \\
\hline
\endhead
\hline
\endfoot

     12.01   &   0   &     11.35   &     -4.83   &       1.8   &       0.7   &      61.5   &    0.0431   &    0.0003  \\
     41.01   &   0   &     11.20   &     -8.62   &       6.1   &       1.9   &      68.5   &    0.0404   &    0.0008  \\
     41.02   &   0   &     11.20   &     -9.77   &      10.1   &       5.8   &      42.1   &    0.0404   &    0.0024  \\
     41.03   &   0   &     11.20   &     -9.61   &       9.4   &       5.2   &      44.9   &    0.0404   &    0.0021  \\
     49.01   &   0   &     13.70   &     -7.31   &       8.1   &       3.4   &      57.7   &    0.0245   &    0.0008  \\
     51.01   &   0   &     13.76   &     -3.75   &       2.9   &       0.1   &      97.9   &    0.0241   &    0.0000  \\
     69.01   &   0   &      9.93   &     -8.45   &       4.1   &       1.5   &      63.7   &    0.0501   &    0.0007  \\
     82.01   &   0   &     11.49   &     -7.15   &       1.3   &       0.2   &      84.7   &    0.0307   &    0.0001  \\
\hline

\end{longtable}
\tablefoot{\\
\tablefoottext{a}{$Type = 0$ for isolated KOIs (no companions within 6 arcsec from the host star) and $Type =1$ for KOIs with at least one companion between 3-6 arcsec (see Table~\ref{tab:photometry} for photometric information about the detected companions).}
}

}

\begin{table*}
\caption{Estimation of the new parameters of the planet candidates orbiting the KOIs with detected companions closer than 3 arcsec. 
}
\scriptsize
\setlength{\extrarowheight}{3pt}
\label{tab:NewParameters}      
\centering                          
\begin{tabular}{c| r c c c c r r}        
\hline\hline                 

        Planet                     & Cat.Depth        & New.Depth           &  cat. $R_p/R_s$          &    new $R_p/R_s$\tablefootmark{a}          &  sec $R_p/R_s$\tablefootmark{b}     &      Cat. $R_p$\tablefootmark{c}   &    New  $R_p $\tablefootmark{d}  \\ 
        candidate                  & ppm              & ppm                 &  $\times 10^{-2}$        & $\times 10^{-2}$          &     $\times 10^{-2}$            & $R_{\oplus}$          & $R_{\oplus}$  \\ \hline

         111.01                    & $496$          & $497.3 \pm 1.7$        & $2.107 \pm 0.020$      & $2.2301 \pm 0.0038$              & $43 \pm 28$                   & $2.14$               & $2.26$  \\
         111.02                    & $455$          & $456.2 \pm 1.6$        & $2.024 \pm 0.023$      & $2.1359 \pm 0.0036$              & $41 \pm 26$                   & $2.05$               & $2.16$  \\
         111.03                    & $598$          & $599.6 \pm 2.0$        & $2.328 \pm 0.026$      & $2.4487 \pm 0.0042$              & $47 \pm 30$                   & $2.36$               & $2.48$  \\
         111.04                     & $56$         & $56.15 \pm 0.19$          & $0.76 \pm 0.11$      & $0.7493 \pm 0.0013$           & $14.5 \pm 9.3$                   & $0.77$               & $0.76$  \\
         191.01                  & $14611$         & $32000 \pm 5800$       & $11.520 \pm 0.051$           & $17.9 \pm 1.6$           & $16.4 \pm 1.2$                  & $11.00$              & $17.10$  \\
         191.02                    & $664$           & $1450 \pm 260$        & $2.426 \pm 0.036$          & $3.82 \pm 0.34$          & $3.49 \pm 0.26$                   & $2.30$               & $3.62$  \\
         191.03                    & $194$             & $425 \pm 77$        & $1.291 \pm 0.043$          & $2.06 \pm 0.19$          & $1.89 \pm 0.14$                   & $1.24$               & $1.98$  \\
         191.04                    & $659$           & $1440 \pm 260$        & $2.402 \pm 0.073$          & $3.80 \pm 0.34$          & $3.48 \pm 0.26$                   & $2.30$               & $3.64$  \\
        1230.01                   & $6998$            & $6998 \pm 17$        & $8.259 \pm 0.018$        & $8.366 \pm 0.010$           & $700 \pm 6100$                  & $37.10$              & $37.58$  \\
        1546.01                  & $14150$           & $19568 \pm 79$       & $10.624 \pm 0.084$       & $13.989 \pm 0.028$         & $22.61 \pm 0.12$                   & $9.50$              & $12.51$  \\
        1812.01                   & $1258$         & $1277.8 \pm 4.1$        & $4.053 \pm 0.065$      & $3.5746 \pm 0.0058$           & $28.5 \pm 2.9$                   & $4.80$               & $4.23$  \\
        2324.01\tablefootmark{e}   & $149$        & $149.39 \pm 0.66$          & $1.10 \pm 0.46$      & $1.2222 \pm 0.0027$              & $24 \pm 21$                   & $0.32$               & $0.36$  \\
        2481.01                    & $793$             & $820 \pm 12$        & $2.750 \pm 0.072$        & $2.865 \pm 0.021$           & $15.3 \pm 3.3$                  & $20.60$              & $21.46$  \\
        3158.01                     & $26$          & $56.8 \pm 10.0$          & $0.47 \pm 0.12$        & $0.753 \pm 0.066$        & $0.707 \pm 0.055$                   & $0.30$               & $0.49$  \\
        3158.02                     & $43$              & $91 \pm 16$          & $0.73 \pm 0.11$        & $0.959 \pm 0.084$        & $0.900 \pm 0.070$                   & $0.47$               & $0.62$  \\
        3158.03                     & $48$             & $103 \pm 18$          & $0.63 \pm 0.12$        & $1.017 \pm 0.089$        & $0.954 \pm 0.074$                   & $0.41$               & $0.66$  \\
        3158.04                     & $52$             & $111 \pm 20$          & $0.65 \pm 0.28$        & $1.055 \pm 0.093$        & $0.990 \pm 0.077$                   & $0.42$               & $0.68$  \\
        3158.05                     & $73$             & $157 \pm 28$          & $0.78 \pm 0.14$          & $1.25 \pm 0.11$        & $1.178 \pm 0.091$                   & $0.51$               & $0.81$  \\
        3263.01                  & $23226$           & $26485 \pm 95$         & $16.88 \pm 0.99$       & $16.274 \pm 0.029$         & $43.44 \pm 0.56$                   & $7.00$               & $6.75$  \\
        3444.01                    & $199$          & $219.6 \pm 6.9$          & $1.59 \pm 0.77$        & $1.482 \pm 0.023$          & $4.64 \pm 0.72$                   & $1.04$               & $0.97$  \\
        3444.02                   & $3285$           & $3620 \pm 110$            & $8.8 \pm 4.9$        & $6.017 \pm 0.095$           & $18.9 \pm 2.9$                   & $5.74$               & $3.93$  \\
        3444.03                     & $96$          & $105.8 \pm 3.3$          & $0.96 \pm 0.49$        & $1.028 \pm 0.016$          & $3.22 \pm 0.50$                   & $0.63$               & $0.67$  \\
        3649.01                 & $110642$      & $1301000 \pm 41000$           & $44.6 \pm 2.7$          & $114.1 \pm 1.8$       & $34.774 \pm 0.050$                  & $65.36$             & $167.18$  \\
        3886.01\tablefootmark{e}   & $441$           & $1350 \pm 300$          & $1.86 \pm 0.12$          & $3.68 \pm 0.41$          & $2.56 \pm 0.14$                  & $25.38$              & $50.31$  \\
        4512.01                   & $3989$            & $5954 \pm 53$          & $5.68 \pm 0.00$        & $7.717 \pm 0.034$       & $10.994 \pm 0.099$                   & $6.19$               & $8.41$  \\

\hline                                   
\end{tabular}

\tablefoot{\\
\tablefoottext{a}{New planet-to-star radii ratio assuming no limb-darkening.}\\
\tablefoottext{b}{Planet-to-star radius assuming that the host is actually the secondary companion detected at less than 3 arcsec. }\\
\tablefoottext{c}{Planet radii calculated by the {\it Kepler} project (http://exoplanetarchive.ipac.caltech.edu)}\\
\tablefoottext{d}{Planet radii assuming the new depth and no limb-darkening. {Please note that this could be the cause that the new derived radii are smaller than catalog radii in some cases.} No error is presented since no error in the stellar radii is given.}\\
\tablefoottext{e}{{According to the UKIRT J-band catalog of the {\it Kepler} field, it remains unclear to us if the detected companions to these KOIs in this paper match some of the targets in the UKIRT catalog.}}
}

\end{table*}

\begin{table*}
\caption{Summary of coincident KOIs in the main high-resolution surveys of the Kepler sample.}             
\setlength{\extrarowheight}{7pt}
\label{tab:comparisson}      
\centering                          
\begin{tabular}{c | c c c c }        
\hline\hline                 %

   &  Lillo-Box  & Adams+12 &  Howell+11  &   Law+13 \\ \hline

Lillo-Box     & 230 & 10 & 12 & 112 \\
Adams+12      &  & 102 & 74 & 66 \\
Howell+11     &  &  & 156  & 85 \\
Law+13        &  &  &  & 714 \\

\hline
\hline

\end{tabular}
\end{table*}
%



\begin{table*}
\caption{Comparison between the improvements in the BSC parameter (in \%) obtained by using the H11 \citep{howell11}, the A12 \citep{adams12}, and our high-resolution images (LB14) for all planet candidates involved (28 in H11 and 27 in A12). In all cases, the BSC has been improved with respect to the speckle images and the A12 study. Note that the common target KOI-0623 to H11 is not presented here because we detected a stellar companion closer than 3 arcsec. The small improvement of the H11 study is mostly due to the reduced field of view, which avoids detection of 1.5-3.0 arcsec companions, where the probability of having a background source is maximum in the 0-3 arcsec range.}             
\scriptsize
\label{tab:h11a12BSC}      
\centering                          
\begin{tabular}{c| c c c c|| c| c c c c}        
\hline\hline                 

Planet&   $P_{BS,0}$   &  \multicolumn{3}{c||}{$P_{BS}$ (\%)}    &     Planet & $P_{BS,0}$  & \multicolumn{3}{c}{$P_{BS}$ (\%)}  \\
candidate & \%     &  H11 & A12  & LB14     & candidate &\%  & H11  & A12 & LB14  \\ \hline

     41.01   &      6.10   &       5.9   &       4.3   &       1.9      111.04   &     13.00   &      12.6   &       8.0   &       9.9   \\
     41.02   &     10.10   &       9.9   &       8.2   &       5.8  &    115.01   &      4.50   &         -   &       2.5   &       1.9  \\
     41.03   &      9.40   &       9.2   &       7.6   &       5.2  &    115.02   &      7.40   &         -   &       5.4   &       4.8  \\
     49.01   &      8.10   &       7.4   &         -   &       3.4  &    115.03   &     14.80   &         -   &      12.8   &      12.2  \\
     69.01   &      4.10   &       4.0   &       2.0   &       1.5  &    196.01   &      4.00   &       2.8   &         -   &       0.4  \\
     82.01   &      1.30   &       1.2   &       0.4   &       0.2  &    245.01   &      0.90   &       0.9   &       0.3   &       0.1  \\
     82.02   &      2.50   &       2.4   &       1.5   &       1.0  &    245.02   &      2.30   &       2.3   &       1.7   &       1.3  \\
     82.03   &      3.10   &       3.0   &       2.1   &       1.6  &    245.03   &      5.30   &       5.3   &       4.7   &       4.2  \\
     82.04   &      4.20   &       4.1   &       3.2   &       2.7  &    245.04   &      4.10   &       4.1   &       3.5   &       3.1  \\
     82.05   &      5.30   &       5.2   &       4.3   &       3.8  &    366.01   &      2.40   &       2.0   &         -   &       0.7  \\
     94.01   &      3.70   &       3.1   &       0.9   &       0.3  &    372.01   &      4.20   &       3.4   &       0.2   &       0.8  \\
     94.02   &     10.60   &      10.0   &       6.5   &       4.1  &    398.01   &      3.80   &       2.5   &         -   &       0.9  \\
     94.03   &      6.80   &       6.2   &       2.7   &       1.2  &    398.02   &      8.50   &       7.2   &         -   &       5.4  \\
     94.04   &     23.80   &      23.2   &      19.7   &      17.2  &    398.03   &     13.50   &      12.2   &         -   &      10.4  \\
    111.01   &      5.40   &       5.1   &       0.9   &       2.3  &    638.01   &     15.20   &         -   &       3.2   &       8.8  \\
    111.02   &      5.60   &       5.3   &       1.0   &       2.6  &    638.02   &     14.60   &         -   &       2.7   &       8.3  \\
    111.03   &      4.90   &       4.6   &       0.7   &       1.9  &            &             &             &             &             \\

\hline                                   
\end{tabular}
\end{table*}

\begin{table*}
\caption{Comparison between the blended source probabilities ($P_{BS}$, in \%) obtained by using the L13 \citep{law13} and our high-resolution images (LB14) for coincident planet candidates (167 in total).}             
\scriptsize
\label{tab:l13BSC}      
\centering                          
\begin{tabular}{r| c c c || r| c c c || r| c c c }        
\hline\hline                 

Planet     & $P_{BS,0}$   &  \multicolumn{2}{c||}{$P_{BS}$ (\%)}   &    Planet & $P_{BS,0}$  & \multicolumn{2}{c||}{$P_{BS}$ (\%)}  &  Planet & $P_{BS,0}$  & \multicolumn{2}{c}{$P_{BS}$ (\%)}              \\ 
candidate  & (\%)   &  L13 & LB14     &    candidate & (\%)  & L13 & LB14   &  candidate & (\%)  & L13 & LB14             \\ \hline

     12.01   &      1.80   &       0.0   &       0.7   &   416.02   &      6.90   &       5.9   &       4.1 &   709.01   &      6.60   &       5.2   &       4.2 \\
     41.01   &      6.10   &       4.6   &       1.9   &   416.03   &     15.50   &      14.5   &      12.7 &   717.01   &      5.40   &       4.6   &       3.9 \\
     41.02   &     10.10   &       8.5   &       5.8   &   431.01   &      4.00   &       2.9   &       2.4 &   717.02   &      8.40   &       7.7   &       6.9 \\
     41.03   &      9.40   &       7.9   &       5.2   &   431.02   &      4.50   &       3.4   &       2.9 &   739.01   &      7.20   &       5.3   &       5.6 \\
     49.01   &      8.10   &       7.1   &       3.4   &   435.01   &      5.80   &       4.9   &       3.3 &   800.01   &     24.20   &      20.6   &      19.3 \\
     69.01   &      4.10   &       3.1   &       1.5   &   435.02   &      2.60   &       1.6   &       0.3 &   800.02   &     23.50   &      20.0   &      18.6 \\
     82.01   &      1.30   &       0.6   &       0.2   &   435.03   &      8.50   &       7.5   &       5.9 &   834.01   &      6.80   &       5.1   &       2.9 \\
     82.02   &      2.50   &       1.8   &       1.0   &   435.04   &     11.50   &      10.5   &       8.9 &   834.02   &     14.30   &      12.6   &      10.4 \\
     82.03   &      3.10   &       2.4   &       1.6   &   435.05   &      7.40   &       6.4   &       4.9 &   834.03   &     17.40   &      15.7   &      13.5 \\
     82.04   &      4.20   &       3.5   &       2.7   &   435.06   &     12.30   &      11.3   &       9.7 &   834.04   &     22.00   &      20.3   &      18.1 \\
     82.05   &      5.30   &       4.6   &       3.8   &   463.01   &     17.90   &      10.7   &       9.0 &   834.05   &     16.20   &      14.5   &      12.3 \\
     94.01   &      3.70   &       1.9   &       0.3   &   478.01   &     10.20   &       2.5   &       6.0 &   884.01   &      6.70   &       5.1   &       3.4 \\
     94.02   &     10.60   &       8.8   &       4.1   &   481.01   &     10.10   &       7.5   &       7.0 &   884.02   &      6.90   &       5.3   &       3.7 \\
     94.03   &      6.80   &       5.0   &       1.2   &   481.02   &     13.80   &      11.1   &      10.6 &   884.03   &     14.10   &      12.5   &      10.8 \\
     94.04   &     23.80   &      22.0   &      17.2   &   481.03   &      9.70   &       7.1   &       6.6 &  1230.01   &      4.10   &      0.02   &       0.2 \\
    111.01   &      5.40   &       4.5   &       2.3   &   528.01   &      7.50   &       6.6   &       5.3 &  1236.01   &      6.30   &       5.5   &       3.8 \\
    111.02   &      5.60   &       4.7   &       2.6   &   528.02   &      6.60   &       5.6   &       4.3 &  1236.02   &      9.30   &       8.5   &       6.8 \\
    111.03   &      4.90   &       4.1   &       1.9   &   528.03   &      7.40   &       6.5   &       5.2 &  1236.03   &      8.40   &       7.6   &       5.9 \\
    111.04   &     13.00   &      12.1   &       9.9   &   534.01   &     19.10   &      14.7   &      14.3 &  1353.01   &      5.20   &       0.8   &       0.2 \\
    115.01   &      4.50   &       3.7   &       1.9   &   534.02   &     24.00   &      19.6   &      19.2 &  1353.02   &     26.00   &      21.6   &      18.1 \\
    115.02   &      7.40   &       6.6   &       4.8   &   561.01   &      5.80   &       4.7   &       4.2 &  1426.01   &      4.40   &       3.3   &       2.7 \\
    115.03   &     14.80   &      13.9   &      12.2   &   564.01   &     21.90   &      19.3   &      15.6 &  1426.02   &      2.20   &       1.1   &       0.6 \\
    139.01   &      3.90   &       3.0   &       0.9   &   564.02   &      9.70   &       7.1   &       3.5 &  1426.03   &      2.10   &       1.0   &       0.6 \\
    139.02   &     15.20   &      14.3   &      12.0   &   564.03   &     35.60   &      33.0   &      29.3 &  1452.01   &      2.70   &       0.4   &       0.1 \\
    149.01   &      6.00   &       5.3   &       2.6   &   567.01   &     12.70   &       9.9   &       8.6 &  1529.01   &      9.80   &       6.3   &       7.8 \\
    152.01   &     13.60   &      10.9   &       3.3   &   567.02   &     14.70   &      11.9   &      10.6 &  1529.02   &     12.90   &       9.4   &      10.9 \\
    152.02   &     25.10   &      22.4   &      14.4   &   567.03   &     13.80   &      10.9   &       9.7 &  1596.01   &     28.40   &      25.5   &      22.7 \\
    152.03   &     26.60   &      24.0   &      16.0   &   571.01   &     28.10   &      14.0   &      20.1 &  1596.02   &     18.00   &      15.1   &      12.3 \\
    152.04   &     31.30   &      28.7   &      20.7   &   571.02   &     26.00   &      11.8   &      17.9 &  1684.01   &      6.40   &       1.9   &       1.5 \\
    156.01   &      9.90   &       5.5   &       6.2   &   571.03   &     32.40   &      18.2   &      24.4 &  1701.01   &     17.20   &      14.7   &      11.9 \\
    156.02   &     12.40   &       8.0   &       8.7   &   571.04   &     27.90   &      13.7   &      19.8 &  1725.01   &      2.40   &       1.6   &       0.8 \\
    156.03   &      6.70   &       2.4   &       3.0   &   571.05   &     33.20   &      19.1   &      25.2 &  1779.01   &      8.60   &       7.4   &       2.7 \\
    191.01   &      6.80   &       3.4   &       0.8   &   579.01   &     10.70   &       8.9   &       7.8 &  1779.02   &     10.80   &       9.6   &       4.6 \\
    191.02   &     28.00   &      24.6   &      21.5   &   579.02   &     10.50   &       8.8   &       7.7 &  1781.01   &      1.70   &       0.4   &       0.3 \\
    191.03   &     43.50   &      40.1   &      36.9   &   611.01   &     11.80   &       6.1   &       2.9 &  1781.02   &      2.80   &       1.5   &       1.3 \\
    191.04   &     28.20   &      24.8   &      21.6   &   624.01   &     10.80   &       8.5   &       5.2 &  1781.03   &      2.10   &       0.7   &       0.6 \\
    209.01   &      2.80   &       1.2   &       0.6   &   624.02   &     11.00   &       8.7   &       5.3 &  1802.01   &      6.90   &       5.6   &       3.5 \\
    209.02   &      4.30   &       2.7   &       1.9   &   624.03   &     17.70   &      15.4   &      12.1 &  1805.01   &      8.70   &       6.6   &       4.2 \\
    211.01   &      3.10   &       1.9   &       0.6   &   625.01   &      5.70   &       4.2   &       3.2 &  1805.02   &     10.40   &       8.3   &       5.9 \\
    238.01   &     23.90   &      19.5   &      16.4   &   632.01   &     13.00   &      11.3   &       9.9 &  1805.03   &     13.70   &      11.7   &       9.2 \\
    238.02   &     39.00   &      34.6   &      31.5   &   638.01   &     15.20   &      11.5   &       8.8 &  1812.01   &      7.30   &       5.3   &       3.1 \\
    330.01   &     16.30   &      10.7   &      11.9   &   638.02   &     14.60   &      11.0   &       8.3 &  1894.01   &     12.80   &       8.3   &       8.2 \\
    330.02   &     29.00   &      23.3   &      24.6   &   640.01   &     21.10   &      19.2   &      10.8 &  1924.01   &      4.70   &       4.4   &       3.2 \\
    339.01   &      7.10   &       6.0   &       5.1   &   650.01   &      8.60   &       6.7   &       7.4 &  1925.01   &      3.50   &       3.0   &       1.7 \\
    339.02   &      7.20   &       6.1   &       5.2   &   654.01   &      8.40   &       7.0   &       6.5 &  2042.01   &     12.60   &      11.3   &       5.8 \\
    339.03   &      7.40   &       6.3   &       5.4   &   654.02   &      9.10   &       7.7   &       7.1 &  2133.01   &      4.30   &       3.7   &       2.3 \\
    345.01   &      3.30   &       2.4   &       2.2   &   659.01   &     15.20   &      13.1   &      10.0 &  2260.01   &      8.90   &       7.5   &       7.1 \\
    349.01   &      5.10   &       2.8   &       3.2   &   664.01   &     15.80   &      13.8   &      11.8 &  2352.01   &      4.60   &       4.0   &       3.2 \\
    366.01   &      2.40   &      0.04   &       0.7   &   664.02   &     21.50   &      19.5   &      17.5 &  2352.02   &      5.10   &       4.5   &       3.7 \\
    372.01   &      4.20   &      0.01   &       0.8   &   664.03   &     21.20   &      19.3   &      17.3 &  2352.03   &      5.20   &       4.7   &       3.8 \\
    385.01   &     19.00   &      16.4   &      16.1   &   676.01   &      5.50   &       0.2   &       2.0 &  2481.01   &     16.60   &      13.1   &      10.0 \\
    386.01   &     16.70   &      14.7   &       9.9   &   676.02   &      7.40   &       1.9   &       3.8 &  2545.01   &      7.70   &       6.7   &       6.4 \\
    386.02   &     18.90   &      16.9   &      12.0   &   682.01   &      5.80   &       2.6   &       0.8 &  2593.01   &      5.60   &       4.8   &       4.3 \\
    388.01   &      8.40   &       7.1   &       6.1   &   684.01   &      4.10   &       3.1   &       2.6 &  2632.01   &      5.50   &       4.7   &       4.8 \\
    393.01   &     23.60   &      22.0   &      17.4   &   686.01   &      3.50   &       0.3   &       0.1 &  2640.01   &      8.80   &       7.3   &       5.0 \\
    416.01   &      6.00   &       5.0   &       3.1   &   695.01   &      5.20   &       4.2   &       3.7 &            &             &             &           \\ 

\hline                                   
\end{tabular}
\end{table*}

\clearpage

\end{document}